\documentclass[draft]{agujournal2019}
\usepackage{url} 
\usepackage{lineno}
\usepackage[inline]{trackchanges} 
\usepackage{soul}
\usepackage{aas_macros}

\draftfalse

\journalname{JGR: Space Physics}

\begin{document}

%
%


\title{Characterization of Magnetic Flux Contents for Flux Transfer Events and its Implications for Flux Rope Formation at the Earth's Magnetopause}
%
%




\authors{Shuo Wang\affil{1}, Ying Zou\affil{2}, Qiang Hu\affil{1,3}, Xueling Shi\affil{4,5}, and Hiroshi Hasegawa\affil{6}}

\affiliation{1}{Center for Space Plasma and Aeronomic Research (CSPAR), The University of Alabama in Huntsville, Huntsville, AL 35805, USA}
\affiliation{2}{Johns Hopkins University Applied Physics Laboratory, Laurel, MD, USA}
\affiliation{3}{Department of Space Science, The University of Alabama in Huntsville, Huntsville, AL 35805, USA}
\affiliation{4}{Department of Electrical and Computer Engineering, Virginia Tech, Blacksburg, VA, USA}
\affiliation{5}{High Altitude Observatory, National Center for Atmospheric Research, Boulder, CO, USA}
\affiliation{6}{Institute of Space and Astronautical Science, Japan Aerospace
Exploration Agency (JAXA), Sagamihara, Japan}




\correspondingauthor{Qiang Hu}{qh0001@uah.edu}




\begin{keypoints}
\item Magnetic flux contents for FTE flux ropes are characterized in terms of the toroidal (axial) and poloidal components
\item The poloidal flux may correspond to the amount of flux ``opened" in the corresponding polar cap region of the ionosphere 
\item {Sequential magnetic reconnection between adjacent field lines injects poloidal flux into the FTE flux rope during its formation process}
\end{keypoints}

%
%

%
%


\begin{abstract}
Flux transfer events (FTEs) are a type of magnetospheric phenomena that exhibit distinctive observational signatures from the in-situ spacecraft measurements across the Earth's magnetopause. They are generally believed to possess a magnetic field configuration of a magnetic flux rope and formed through magnetic reconnection at the dayside magnetopause, sometimes accompanied with enhanced plasma convection in the ionosphere. We examine two FTE events under the condition of southward interplanetary magnetic field (IMF)  with a dawn-dusk component at the magnetopause by applying the Grad-Shafranov (GS) reconstruction method to the in-situ measurements by the Magnetospheric Multiscale (MMS) spacecraft to derive the magnetic flux contents associated with the FTE flux ropes. In particular, given a cylindrical magnetic flux rope configuration derived from the GS reconstruction, the magnetic flux content can be characterized by both the toroidal (axial) and poloidal fluxes.  We then estimate the amount of magnetic flux (i.e., the reconnection flux) encompassed by the area ``opened" in the ionosphere, based on the ground-based Super Dual Auroral Radar Network (SuperDARN) observations. We find that for event 1, the FTE flux rope is oriented in the approximate dawn-dusk direction, and the amount of its poloidal magnetic flux agrees with the corresponding reconnection flux. For event 2, the agreement among the estimates of the magnetic fluxes is uncertain. We provide a detailed description about our interpretation for the topological features of the FTE flux ropes, based on a formation scenario of sequential magnetic field reconnection between adjacent field lines, consistent with our results.
\end{abstract}

\section*{Plain Language Summary}
The outer boundary of the Earth's own magnetic field extends into space and is shaped by the constant outflow of ionized particles from the Sun, i.e., the so-called solar wind, into a bullet shape. The blunt side facing the Sun is called the dayside magnetopause where the Sun's magnetic field carried along by the solar wind interacts with the Earth's magnetic field. Under the condition of the Sun's magnetic field possessing a southward component, the interaction becomes more intense and energetic, often leading to a continuous change of topology/connectivity between the two fields. Such a process, dubbed magnetic reconnection,  is also accompanied with enhanced particle motion, of which signatures can manifest in the in-situ spacecraft measurements. Correspondingly such enhanced disturbances may map nearly simultaneously along the Earth's magnetic field lines onto the Earth's upper atmosphere and observed by the ground-based radars. By analyzing and correlating these observations at different but inter-connected sites, we carry out a study to characterize and relate the physical quantity of magnetic flux accumulated through the reconnection process. We also illustrate in detail the formation of one type of commonly associated magnetic field structure at the dayside magnetopause. 

\section{Introduction} \label{sec:intro} 
Flux transfer events (FTEs) are recognized as signatures of intermittent magnetic reconnection from in-situ spacecraft measurements during the crossings of the Earth's magnetopause \cite{1978SSRv...22..681R,1990GMS....58..455E,2022SSRv..218...40Z}. They generally possess the signatures of bipolar magnetic field component in the direction normal to the local plane of the magnetopause current sheet, and sometimes elevated magnetic field magnitude and bursty plasma flows. The polarity of this normal field component follows certain pattern with respect to the locations of their occurrence, owing to the nominal field directions across the magnetopause.  They have typical duration around 1 minute and occur most often and sometimes repeatedly under the southward interplanetary magnetic field (IMF) or magnetosheath magnetic field (shocked IMF) conditions. Additional plasma and particle signatures support the generation mechanism of magnetic reconnection and the magnetic field topology of a magnetic flux rope \cite{1990GMS....58..455E,2006AnGeo..24..381R,2012MEEP....1...71H,2021JGRA..12629388G} for FTEs. Interestingly, in \citeA{1990GMS....58..455E}, it was indicated that ``What Russell and Elphic [1978] suggested, in effect, was a magnetopause analog to solar flares". For solar flares, magnetic reconnection always plays a critical role, often leading to the formation of magnetic flux ropes on the Sun \cite{Forbes2000,2006SSRv..123..251F,2011LRSP....8....1C}. We will further digress on this aspect and offer our view on this analogy with greater details in Section~\ref{sec:interp}.
In this aspect for FTEs at the Earth's magnetopause, 
a flux rope topology is conceived to be formed through the process of single or multiple X line reconnection  \cite{2017JGRA..12212310F,1985GeoRL..12..105L,2010GL044219HH}. For the latter, the flux rope may possess a more pronounced non-vanishing axial field component, thus exhibiting a configuration of helical magnetic field lines \cite<e.g.,>[]{1990GMS....58..515F}.

Magnetic flux ropes are a common and important type of structures occurring across space plasma regimes and magnetic reconnection is believed to play a major role in the formation of flux ropes \cite{1990GMS....58.....R}. They are observed on the Sun, in the interplanetary space, at the Earth's magnetopause, as well as in the magnetotail, from both in-situ and remote-sensing  observations. 
In particular, for the in-situ spacecraft measurements, the Grad-Shafranov (GS) reconstruction method has been widely applied to derive the configuration of magnetic flux ropes in various space plasma regimes and with a wide range of scale sizes, including FTEs at the Earth's magnetopause \cite{2012MEEP....1...71H,Hu2017GSreview}. In these applications to FTEs, the method has been validated by using multi-spacecraft measurements and the results were interpreted in the context of approximately two and a half dimensional (2$\frac{1}{2}$-D) flux ropes formed through magnetic reconnection \cite{Hasegawa2004,angeo-24-603-2006}. { In \citeA{angeo-24-603-2006}, two groups of possibly recurring FTEs were examined by the optimal GS reconstruction technique by employing multiple Cluster spacecraft datasets, which enabled the most accurate characterization of the FTE flux rope configurations. It was found that the cross section size of an FTE flux rope can reach the order of $\sim$1 Earth radius ($R_E$), and they all possess a strong core (axial) field. The results indicated consistency with the usual single-spacecraft based GS reconstruction results. In addition, the flux contents of the FTE flux ropes were also quantified, in terms of the axial flux and the ``total transverse magnetic flux" (equivalent to the poloidal flux as we refer later). Most noteworthily, those authors were able to derive the reconnection rate (in the order of $<0.1$ in normalized unit) for the FTE formation based on the realization that the ``total reconnected flux" is equal to the poloidal flux of the flux rope. Part of their analysis result is to be cited in Section~\ref{sec:results} for reference. Another inspiration is the series of recent works by \citeA{2021GL096583Zou,2021JA029117Zou,2017GL075765Zou}, albeit not directly addressing FTEs. Those authors  have carried out detailed and correlated analysis of both in-situ spacecraft measurements and ground-based  observations under the ``space-ground conjunction". The dayside magnetopause reconnection processes were studied especially in terms of the reconnection rates at the conjugate sites of reconnected field lines with one end connecting to the ionosphere.  In the present study, we also seek out events of such conjunctions with correlated in-situ and ground-based observations, but focus on the utilization of single-spacecraft dataset to derive the critical parameters for FTE flux ropes in order to correlate with the associated physical quantities derived from the corresponding radar observations in the ionosphere. We intend to further elucidate the process of magnetic reconnection at the magnetopause, leading to the formation of FTE flux rope in detail, from a topological point of view.}

FTEs have also been studied by using optical/radar observations in addition to in-situ spacecraft measurements. Poleward Moving Auroral Forms (PMAFs) are a type of auroral structure that is observed remotely and occurs  in the ionosphere \cite{1975P&SS...23..269V,1986JGR....9110063S}. PMAFs are caused by the reconnection of magnetic field lines in the magnetosphere and the magnetosheath or the boundary layers across the magnetopause, which process forms FTEs. The ionospheric signatures of FTEs through mapped field lines from the magnetopause to the ionosphere can be observed optically as PMAFs. There are corresponding signatures occurring  at the footprints of newly opened magnetic field lines and are characterized by a poleward motion of the associated plasma structures \cite{2019JA027674}. Such a connection was made by using both in-situ spacecraft measurements of an FTE at the magnetopause and the corresponding radar and camera observations in the ionosphere with enhanced plasma convection and auroral structures near the conjugate sites that map to the FTE location  \cite{1990GeoRL..17.2241E,2003AnGeo..21.1807W}.  The Super Dual Auroral Radar Network (SuperDARN) observations \cite{1995SSRv...71..761G,2007SGeo...28...33C,2019PEPS....6...27N} have been used to analyze the motion and estimate the area ``opened" by such a magnetic reconnection process. For example, some previous studies \cite{1990JGR....9517113L,2000JGR...10515741M} inferred the latitudinal and longitudinal extents of ``opened" magnetic field region using radar and auroral observations. 
In turn, a connection can be made between the FTE formation at the magnetopause and the corresponding signatures in the ionosphere. In particular, certain amount of flux for the reconnected field (hence the reconnection flux) can be estimated by using the radar observations to provide a quantitative characterization that can be compared with the corresponding FTE fluxes \cite{2004AnGeo..22..141M,2004GeoRL..31.9809M,2005AnGeo..23.2657O,2017JGRA..12212310F}.  It was summarized by \citeA{2017JGRA..12212310F} that the  range of magnetic flux contents for the conjugate FTE events is approximately between 1 and 77 MWb. 

In this study, we follow the overall approach of  \citeA{2017JGRA..12212310F}, especially for analyzing the radar observations, but instead applying the GS reconstruction method to the in-situ spacecraft measurements of FTEs, in order to estimate the magnetic flux contents associated with FTE formation processes. We describe the data source and methods employed in Section~\ref{sec:methods}. The results for two events from the analysis of both in-situ spacecraft measurements and the associated radar observations are presented in Section~\ref{sec:results}. Based on these analysis results, we offer an interpretation for the FTE formation process at the magnetopause in Section~\ref{sec:interp} solely from the viewpoint of topological change of magnetic fields. Finally we conclude and discuss the implications and uncertainties associated with this analysis.

\section{Data and Methods}\label{sec:methods}
Following \citeA{2017JGRA..12212310F}, we utilize both in-situ spacecraft measurements, primarily from the Magnetospheric Multiscale (MMS) spacecraft, at the magnetopause and the corresponding SuperDARN observations in the ionosphere to carry out the quantitative analysis of the magnetic flux contents associated with the FTE flux ropes and the reconnection flux ``opened" in the polar region of the ionosphere.
The  MMS mission is a constellation of four spacecraft to study the Earth's magnetosphere and the important process of magnetic reconnection through in-situ measurements of magnetic field and particle populations. The magnetic field data are gathered through the use of a fluxgate magnetometer \cite{2016SSRv..199..189R}, with a sampling rate of 128 Hz. The Fast Plasma Investigation (FPI) instrument \cite{2016SSRv..199..331P} is used to obtain the ion and electron distribution functions and to derive their associated moments. Only data obtained in burst mode are utilized in this study and  are in the Geocentric Solar Magnetospheric (GSM) coordinate system from the MMS1 spacecraft. 

SuperDARN  is a global network of scientific radars located in both the Northern and Southern Hemispheres. The SuperDARN radar data are used to map high-latitude plasma convection and to display back scatter power and Doppler velocity for a selected beam along a particular line of sight in this study. The convection map is generated from the improved model of \citeA{2018JA025280} (TS18 model). The technique uses data from all the SuperDARN stations in one hemisphere and data from a statistical model for regions without real-time radar observations. We follow closely the procedures given by \citeA{2017JGRA..12212310F} for quantifying the amount of flux in the polar cap region ``opened" by the magnetic reconnection associated with the corresponding FTE formation at the magnetopause. 
Namely, the longitudinal and latitudinal extents of the area ``opened" are estimated by the extents of the enhanced plasma convection velocities and the poleward propagation of the enhanced back scatter power, respectively. The expansion of the enhanced radar scatter power is considered equivalent to the signatures of PMAFs in our analysis.

To characterize the magnetic flux contents of an FTE flux rope, we take a different and unique approach by employing the Grad-Shafranov (GS) reconstruction method based on in-situ data. The GS method has been applied to examine the magnetic field structures of FTEs at the Earth's magnetopause \cite{Sonnerup2004,Hasegawa2004,angeo-24-603-2006,2012MEEP....1...71H,Hu2017GSreview}, in the form of a cylindrical flux rope configuration composed of nested flux surfaces with arbitrary (2D) cross sections. Through this approach, the critical parameters characterizing a flux rope structure can be derived quantitatively, including the magnetic flux contents. 

The GS reconstruction method employs the GS equation in a Cartesian coordinates which governs the magnetic flux function $A(x,y)$ in a 2D geometry (i.e., $\partial/\partial z=0$),
\begin{equation}
\frac{\partial^2{A}}{\partial{x}^2}+\frac{\partial^2{A}}{\partial{y}^2}=-\mu_0\frac{dP_t(A)}{dA}. \label{eq:GS}
\end{equation}
Here, due to the invariance along the $z$ dimension ($z$ being the cylindrical axis), the magnetic field components are determined by the scalar magnetic flux function, via, $B_x = \partial A/\partial y$, $B_y = -\partial A/\partial x$, and $B_z=B_z(A)\ne 0$. On the right-hand side, the total derivative with respect to $A$ involves the so-called transverse pressure 
$P_t(A)=p(A)+B_z^2(A)/{2\mu_0}$, which is a single-variable function of $A$ and the sum of the plasma pressure and the axial magnetic pressure. Therefore a 
solution $A(x,y)$ to the GS equation fully characterizes a cylindrical magnetic field configuration with all three field components including the non-vanishing axial component known over the cross-section plane perpendicular to the $z$ axis. 

The GS reconstruction procedures proceed by integrating the flux function from the initial spacecraft path at $y=0$ where the initial values are known from the spacecraft measurements once an optimal $z$ axis orientation is determined \cite{2002JGRAHu} together with a proper frame of reference in which the structure is in approximate magnetohydrostatic equilibrium. The reference frame is chosen as the deHoffmann-Teller (HT) frame with the frame velocity $\mathbf{V}_{HT}$ which is determined from the magnetic field and plasma velocity measurements \cite{2008ISSIR...8...65P,1998ISSIRK}. The quality of the HT frame is assessed by a correlation coefficient $cc_{HT}$ (1 being ideal) and the Wal\'en test slope (0 being ideal). The latter evaluates the relative magnitude of the remaining plasma flow in the HT frame with respect to the local Alfv\'en speed. One essential step involves an analytic function fitting to the quantities $P_t$ versus $A$ in order to make the right-hand side of the GS equation~(\ref{eq:GS}) explicitly known, i.e., by obtaining an analytic functional form $P_t(A)$ through curve fitting. The same procedure is applied for obtaining $B_z(A)$. The end result is a 2D array of $A(x,y)$ over a rectangular domain, together with the distribution of $B_z$. Thus all three components of the magnetic field are obtained as functions of $(x,y)$. In addition to a number of standard output quantities, the solution can be specifically utilized to calculate the axial (toroidal) magnetic flux $\Phi_z$ and the poloidal magnetic flux $\Phi_p$ of a flux rope configuration in a precise way \cite{Qiu2007,2014ApJH}:
\begin{equation}
    \Phi_z=\int_S B_z dS,\label{eq:Phiz}
\end{equation}
and
\begin{equation}
    \Phi_p=|A_m - A_b|\cdot L=\phi_p\cdot L.\label{eq:Phip}
\end{equation}
Here an area $S$ is chosen over the cross section plane, within which the axial flux can be summed up for the central region of a magnetic flux rope. A physical choice of the boundary for such an area is $A=A_b$ based on the $P_t(A)$ or $B_z(A)$ fitting, which indicates that  the solution within this boundary (as highlighted by the white contour in Figure~\ref{fig:GSmap1}a) is judged to satisfy  the GS equation under certain threshold conditions (e.g., for a fitting residue of $P_t(A)$, $R_f\ll 1$). In this way, a boundary is specified by a flux surface that has an arbitrary cross section shape resulting in  a truly 2D structure, as we will illustrate in the following event studies, based on in-situ spacecraft data. 

More straightforwardly, owing to the definition of the flux function for a 2D geometry, the flux function $A$ itself directly characterizes distinct flux surfaces. The difference in $A$ between a pair of such distinct surfaces represents the amount of unit poloidal flux $\phi_p$ enclosed by a rectangular area intercepting and bounded by these two surfaces with a unit axial length in the $z$ dimension. Therefore for a flux rope of axial length $L$ and a boundary at $A=A_b$, the amount of poloidal flux is given by equation~(\ref{eq:Phip}) where the flux function value at the center of the flux rope (corresponding to the extremum in $A$ inside the flux rope boundary) is denoted $A_m$.

\section{Analysis Results}\label{sec:results}
\begin{table}
 \caption{Event parameters for the two FTEs and the corresponding GS reconstructions.}\label{tbl:1}
 \centering
 \begin{tabular}{l c c }
 \hline
  Event & 1 & 2\\
  Date & 27 November 2016& 19 December 2016  \\
  Time interval (UT) & 08:39:08 - 08:40:05 & 09:15:40 - 09:17:46 \\
  MMS1 location (GSM) [$R_{E}$] & (10.3, 3.6, -1.4) & (11.8, 1.8, 0.3)    \\
  Optimal $z$ axis (GSM) & (-0.161, 0.825, 0.542)& (0.057, 0.064, -0.996)  \\
  HT frame velocity [km/s] & (-93, 212, -114)& (-22, 92, -48)  \\
  Wal\'en test slope & -0.19 &  -0.28 \\
  $cc_{HT}$ & 0.89 &  0.72 \\
  Chirality & right-handed & left-handed \\
  Axial flux $\Phi_z$ [MWb]  & 3.4 & 5.3 \\
  Unit poloridal flux $\phi_p$ [MWb/$R_{E}$] & 0.684& 0.763  \\
 \hline
 \end{tabular}
 \end{table}

 \begin{figure}
\includegraphics[width=33pc]{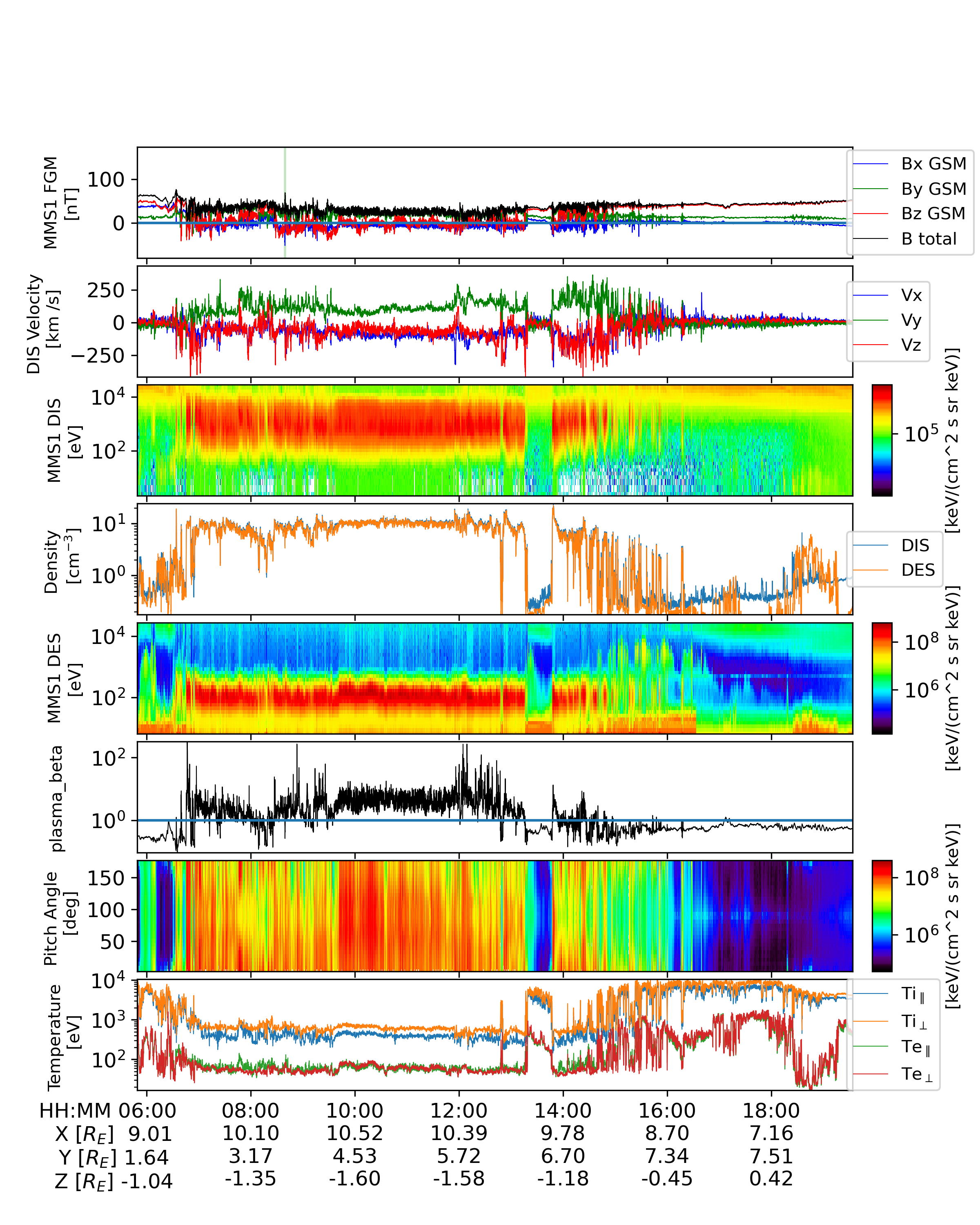}
\caption{Times-series measurements from the MMS1 spacecraft for event 1 on 27 November 2016. From the top to the bottom panels are the GSM components of the  magnetic field and the field magnitude, the ion velocity from the dual ion  spectrometers (DIS), the ion energy spectrogram, the number density from DIS and the dual electron spectrometers (DES), the electron energy spectrogram, the  plasma $\beta$, the electron pitch angle distribution (ePAD) for the 0.2-2 keV electrons,  and the perpendicular and parallel temperature for ions and electrons. See the legends and labels for details. The MMS1 spacecraft locations in the GSM coordinates are also listed beneath the time tick labels.  The light green vertical lines in the top panel mark the time interval of the FTE flux rope for event 1.}\label{fig:MMSdata1}
\end{figure}

\begin{figure}
\includegraphics[width=33pc]{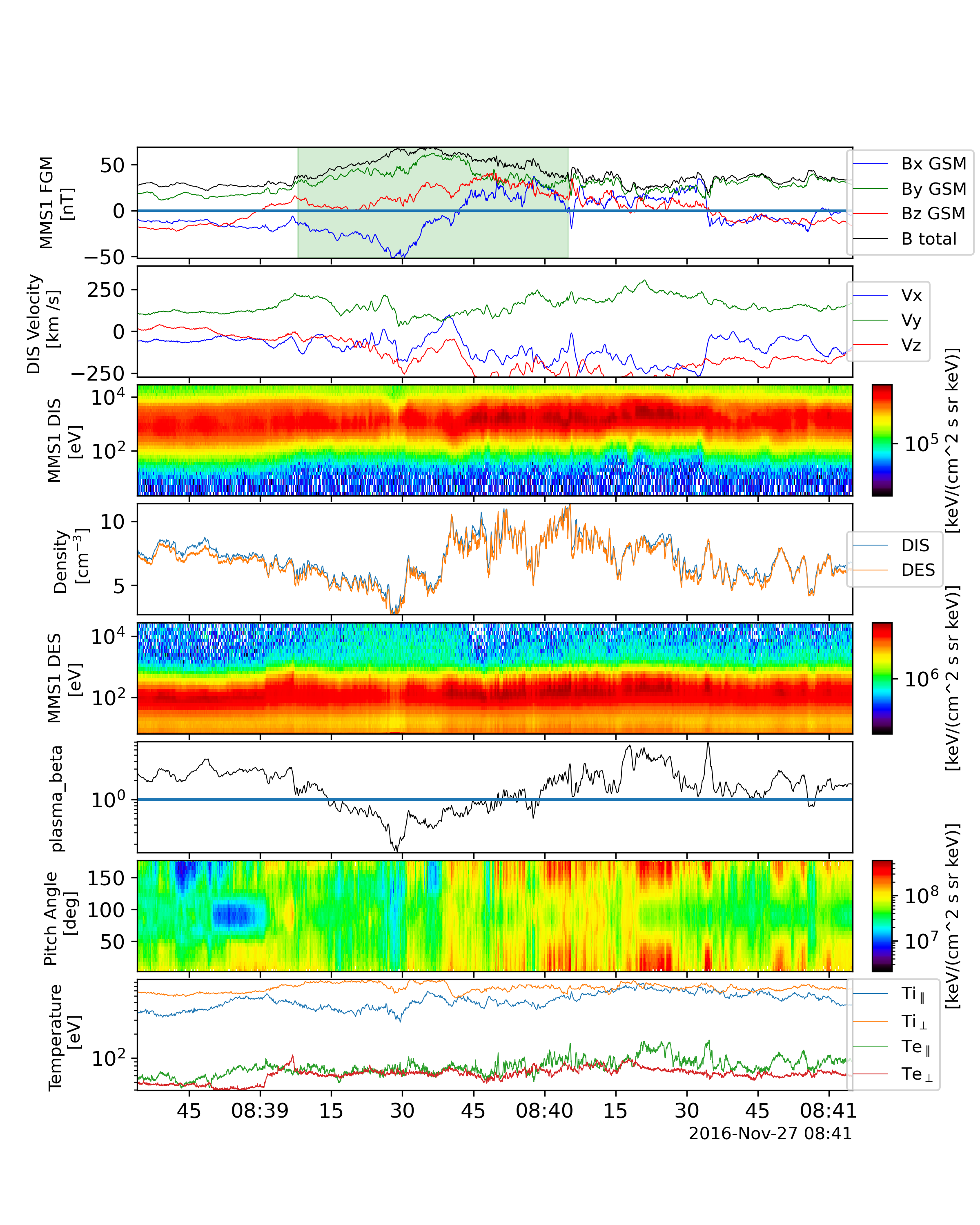}
\caption{The same time-series stack plot as Figure~\ref{fig:MMSdata1} but for a much shorter time period surrounding the FTE interval, which is  marked in the top panel by the light green shaded area. \label{fig:short1}}
\end{figure}
\subsection{Event 1: 27 November 2016}\label{sec:event1analysis}
Figure~\ref{fig:MMSdata1} shows an overview plot of the in-situ measurements from MMS1 over a $\sim$ 12 hour time period on 27 November 2016. It shows mostly typical magnetosheath conditions but with a final transition into the magnetosphere around 16:00 UT. 
The FTE interval, based on the event list provided by \citeA{2020GeoRL..4786726F}, is marked in the top panel, which has a duration $\sim$ 1 minute. The zoomed-in FTE interval is shown in Figure~\ref{fig:short1} with the same set of panels, but for a much shorter time period surrounding the interval selected for the GS reconstruction. Such an interval, as marked, shows clear signatures for a possible magnetic flux rope structure. The magnetic field magnitude is relatively stronger than the surrounding field, and two field components, $B_Z$ and $B_X$, show gradual rotations, while the $B_Y$ component is unipolar and is significant in magnitude. The plasma $\beta$ value decreases below 1.0 near the central portion of the interval. These magnetic field signatures hint at a magnetic flux rope configuration. The structure is likely oriented horizontally in the dawn-dusk direction at the magnetopause, given the spacecraft location and the relative spacecraft path across such a structure along the $-\mathbf{V}_{HT}$ direction, as listed in Table~\ref{tbl:1}. 
{Figure~\ref{fig:short1} shows southward and duskward enhancements of the ion
velocity (signature of reconnection jets, the second panel). The HT velocity is southward
and duskward (consistent with the expected motion of an FTE flux rope), and
the electron temperatures are higher than in the surrounding
magnetosheath region. In the pristine magnetosheath, the electron
perpendicular temperature tends to be higher than the parallel
temperature (not always though; \citeA{Phan93JA02444}), but for the event 1 interval, the parallel
temperature tends to be higher, which is a signature of reconnected
field lines. The ePAD plot also shows enhanced bi-directional field-aligned streaming magnetosheath electrons heated by
the magnetopause reconnection \cite{2010GL044219HH}.  }

\begin{figure}
\centering
\includegraphics[width=.55\textwidth]{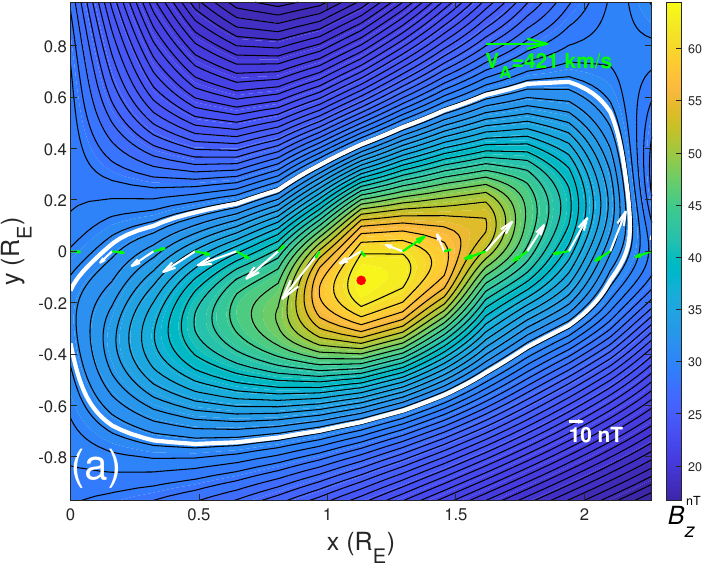}
\includegraphics[width=.44\textwidth]{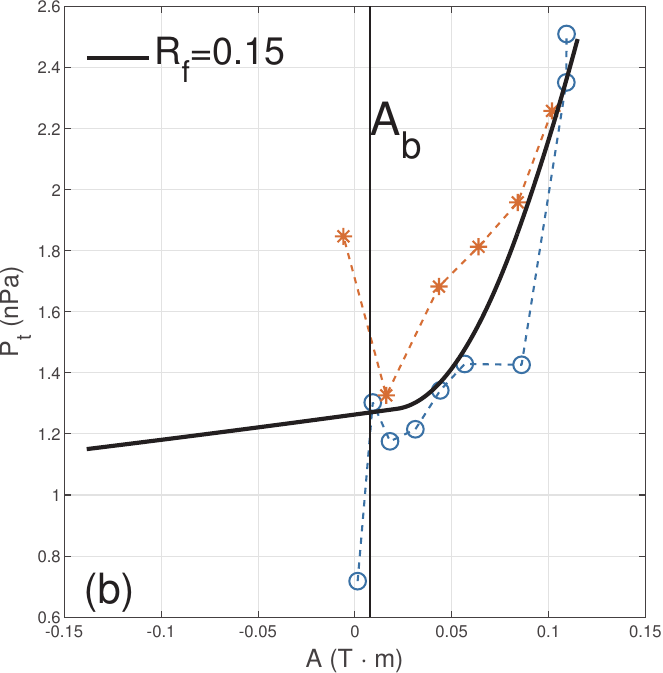}
\caption{The GS reconstruction result for the event 1 FTE interval based on MMS1 spacecraft measurements. (a) The cross section of the magnetic field structure on a plane perpendicular to the $z$ axis reconstructed from the spacecraft measurements along its path ($y=0$). The black contours are the contours of the flux function $A$ (also the transverse field lines on the plane), and the color represents the $B_z$ distribution as indicated by the color bar. The red dot marks the location of the maximum $B_z$ value. The white (green) arrows along $y=0$ indicate the measured transverse magnetic field (remaining plasma flow velocity in the HT frame). Reference vectors are given near the bottom  and top right corners, as denoted by a magnitude of 10 nT for the magnetic field and the average local Alfv\'en speed $V_A$ for the velocity, respectively. The length of the reference vector for the magnetic field is equivalent to 0.20$V_A$.  (b) The corresponding $P_t$ versus $A$ measurements along the spacecraft path (circle and star symbols) and the associated fitting curve $P_t(A)$ in black. A fitting residue, $R_f$, is calculated to indicate the quality of fitting  as denoted. The vertical line marks the choice of a particular flux function value $A_b$ which defines a boundary of the flux rope structure as highlighted by the white thick contour where $A=A_b$ in (a). }\label{fig:GSmap1}
\end{figure}

The GS reconstruction is carried out, yielding a $z$ axis orientation mostly along the GSM-Y direction. A cross-section map of the magnetic field configuration is presented in Figure~\ref{fig:GSmap1}a, together with the corresponding $P_t(A)$ plot in (b). The flux rope configuration is seen as represented by the closed contours of the flux function $A(x,y)$ (i.e., equivalent to nested flux surfaces in this view down the $z$ axis), bounded by the white contour at which $A=A_b$. Such a boundary as highlighted indicates that within this cylindrical ($2\frac{1}{2}$ D) flux surface, the cylindrical flux rope configuration with nested flux surfaces is more reliably reconstructed because those  surfaces are crossed by the spacecraft along its path with actual measurements returned as the data points given in Figure~\ref{fig:GSmap1}b. Therefore the reconstruction result obtained within this flux rope boundary is mostly consistent with the spacecraft measurements for this event, as judged in part by an acceptable fitting residue value $R_f=0.15$ (for the corresponding fitting of $B_z(A)$, $R_f=0.08$). The flux rope possesses right-handed chirality (positive sign of magnetic helicity). The magnetic flux contents are estimated based on the GS reconstruction result and are given in Table~\ref{tbl:1}. The axial flux is a summation of the axial flux element over an area enclosed by the flux rope boundary, within which $A>A_b$ for this event. The total poloidal flux is subject to the determination of the axial length, $L$, of the flux rope along the $z$ dimension. It is determined in coordination with the corresponding radar observations in the ionosphere as to be described below. 

\begin{figure}
\includegraphics[width=\textwidth]{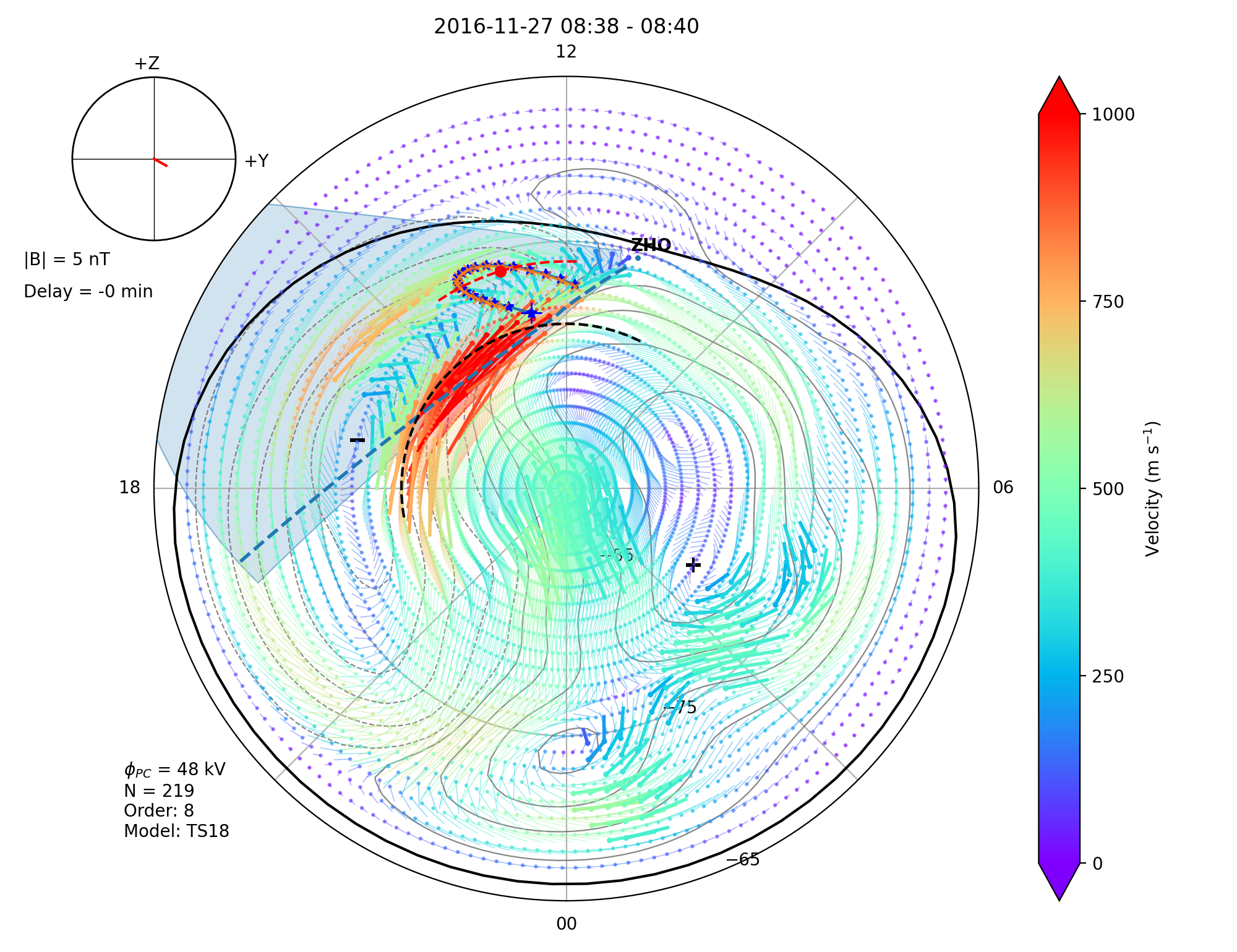}
\caption{Convection map for event 1 over a set of regular magnetic longitudinal (in magnetic local time, MLT) and latitudinal (in degrees, MLAT) grid points (dots in the background) of the Southern Hemisphere centered around the  pole derived from the SuperDARN observations. Thick vectors are velocities as measured by the ground-based radar network, while thin vectors are fitted values. All  point away from the dots and their magnitudes are color coded according to the colorbar to the right. The field of view of the Zhongshan station (ZHO) is shown as a light blue fan-shaped sector, with one beam position marked by a dashed blue line across a region of enhanced poleward flow. The red dot represents the conjugate footpoint of the magnetic field line connected to the MMS1 spacecraft position at the magnetopause.  The Heppnar-Maynard Boundary is plotted as a black solid contour. Two sets of points along the axial direction of the reconstructed FTE flux rope structure at the magnetopause are mapped to the ionosphere as marked by the blue stars and the orange crosses near the red dot. Each point along the axis of the reconstructed flux rope is 1 $R_{E}$ apart from its neighboring points. The end point marked by the blue plus sign corresponds to the most dusk-ward  end point (toward the +GSM-Y direction) along the axis of the flux rope at the magnetopause. The dashed black curve denotes the range of longitudes across the enhanced flow region along which the velocity measurements are taken for determining a longitudinal extent of the ``opened" magnetic flux region which is marked by the red dashed curve along the same latitude as the mapped MMS1 spacecraft position. The nominal IMF condition  is shown by a projection down the +X direction in the top-left corner where the average magnetic field is given by the red line with the circle indicating a field magnitude of 5 nT.   \label{fig:convmap1}}
\end{figure}

\begin{figure}
\centering
\includegraphics[width=.49\textwidth]{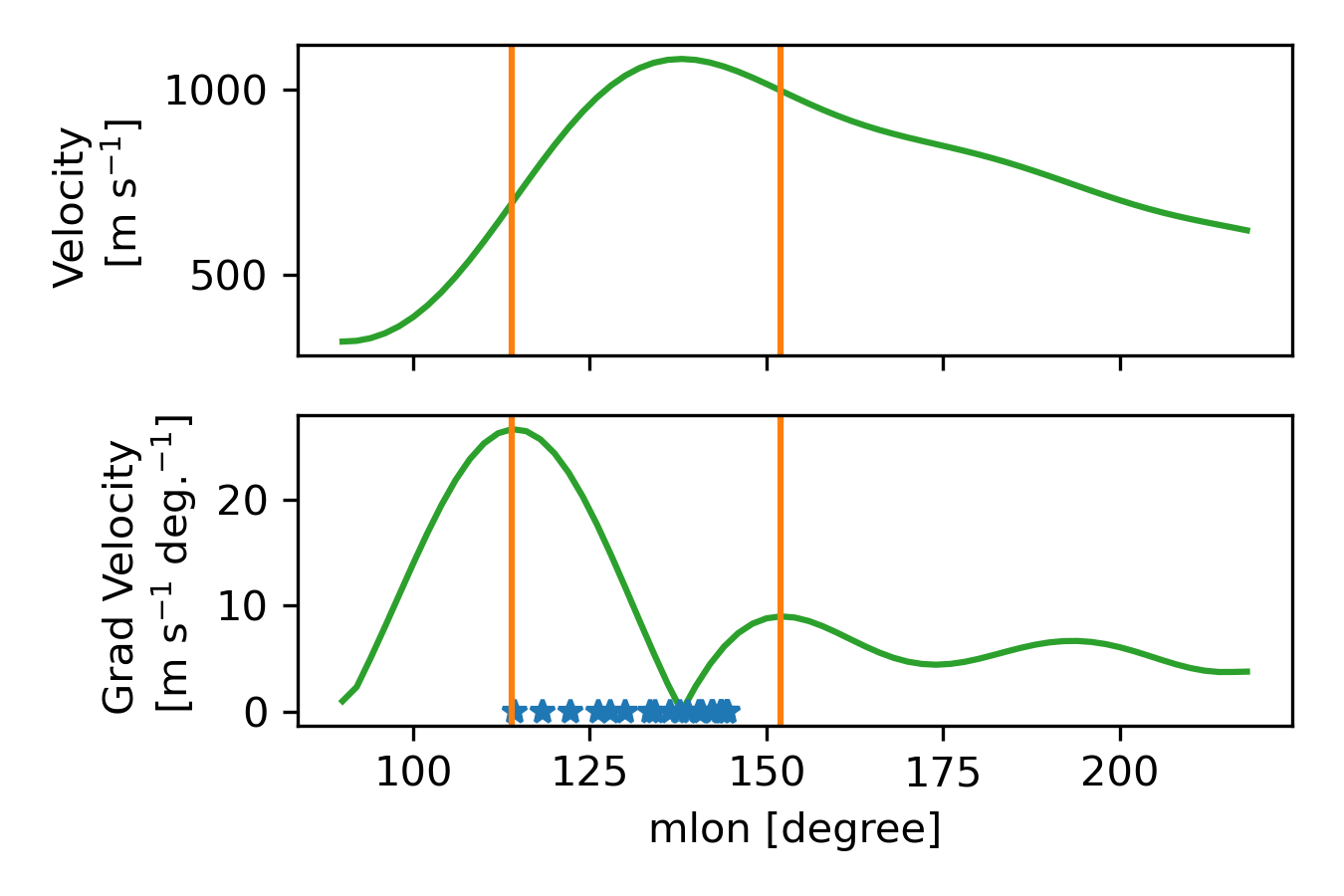}
\includegraphics[width=.5\textwidth]{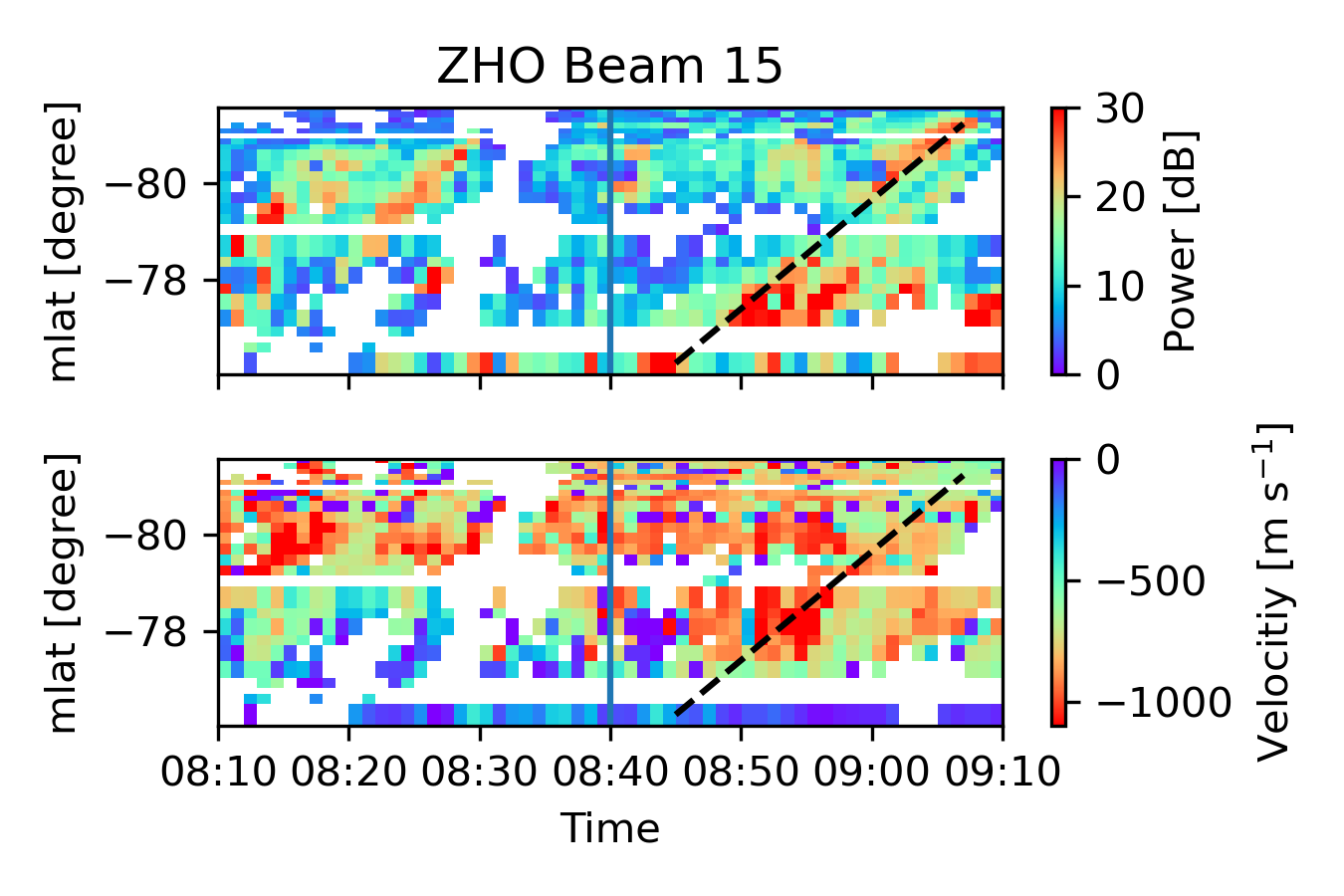}
\hspace{.1\textwidth} (a) \hspace{.4\textwidth} (b)\\
\caption{Analysis of the longitudinal and latitudinal extents of the opened flux region in the ionosphere for event 1. (a) Upper panel: the magnitude of  velocities along the dashed black curve in Figure~\ref{fig:convmap1}. Lower panel: the absolute value of the gradient of the velocities. Vertical lines mark the peaks in the magnitude of the gradient. The range between the vertical lines is taken as the longitudinal extent of the ``opened" flux region. Blue stars belong to the same set of symbols marked in Figure~\ref{fig:convmap1}, but are lined up within the marked range of longitudes only. (b) Radar observation from Beam 15 of the Zhongshan station (ZHO) as a function of time and magnetic latitudes. Upper panel: the radar back scatter power. Dashed black line shows a guideline for the propagating feature with enhanced scattering around the time of the FTE interval. Lower panel: the line-of-sight velocity. The vertical blue lines mark the beginning time of the FTE interval observed by the MMS1 spacecraft.  \label{fig:extents}}
\end{figure}

The analysis of the corresponding radar observations is  carried out following the general procedures described in \citeA{2017JGRA..12212310F}, under the assumption that the signatures of FTE formation, in the form of reconnected field line footpoints motion, map to the polar cap region of the ionosphere nearly simultaneously. Figure~\ref{fig:convmap1} shows the corresponding 
convection map in the Southern Hemisphere above the 65$^\circ$ magnetic latitude in its usual format. Ionospheric flows between 08:38 and 08:40 UT on 27 November 2016 are plotted on the altitude adjusted corrected geomagnetic (AACGM) coordinates \cite{2014JA020264Shepherd}. The plasma convection  pattern with two cells in the Earth's ionosphere is consistent with the southward directed IMF. The footprint of MMS1 is traced along the magnetic field line according to the \citeA{1996ESASP.389..181T} model down to the ionosphere and is marked by the red dot which is at (13.0 MLT, -76.2$^{\circ}$ MLAT). Nearby a region with enhanced flow on the dayside is observed in the post-noon sector at a latitude around -80$^{\circ}$  and is  within the ZHO coverage. The dashed black curve marks the longitudinal range of the enhanced flow region at this latitude. To determine the longitudinal extent of the ``opened" flux region, we plot in Figure~\ref{fig:extents}a the magnitudes of the flow velocity and its gradient along this particular latitude. The extent is taken as the range between the two vertical lines, about 38$^{\circ}$ in longitude, and is marked by the red dashed curve at the same latitude as the red dot in Figure~\ref{fig:convmap1}.

The back scatter power and line-of-sight velocity measured by beam 15 of ZHO are displayed in Figure~\ref{fig:extents}b.  The latitudinal expansion equivalent to the  PMAFs is shown by the dashed line. There is a time difference of 4 minutes between the FTE flux rope interval and the onset of PMAFs. Depending on the propagation time between the FTE location at the magnetopause and the conjugate field-line footpoints in the ionosphere, it is possible to have a time difference of a few minutes \cite{2003AnGeo..21.1807W}. The PMAFs start at -76$^{\circ}$ latitude, and move poleward to -81$^{\circ}$ latitude. The line-of-sight velocities of the PMAFs reach about 900 m/s away from the ZHO station.
The PMAFs  are thus observed to have propagated by 5$^{\circ}$ of magnetic latitudes into the polar cap, corresponding to a poleward distance of $\sim$500 km. 
The area of the polar cap opened by the corresponding FTE formation via magnetic reconnection  is the product of the linear lengths of the above estimated latitudinal and longitudinal extents, which is approximately 0.56 Mm$^2$. The radial ionospheric magnetic field strength is 5 $\times$ 10$^{-5}$ T.
 Following \citeA{2017JGRA..12212310F}, we assume that the uncertainties in the MLT extent and in the latitudinal direction are  $\pm$1 h  and $\pm$2$^{\circ}$, respectively. The reconnection flux $\Phi_R$ calculated using the radar data is 28 $\pm$ 16 MWb for this event.

One unique advantage of combining the two sets of observations at the magnetopause and in the ionosphere is to help refine the analysis of the flux rope configuration by addressing the uncertainty associated with determining the axial length of a cylindrical flux rope for the FTE event. {Similar to \citeA{2017JGRA..12212310F}, by establishing a mapping between the extent of the ``opened" region in the ionosphere to the magnetopause, a finite axial length can be determined. However our approach is different in that we start the mapping from the magnetopause based on the GS reconstruction result by selecting a series of points, separated by 1 $R_E$ in this case, extending along the flux rope axial direction from the locations  on the spacecraft path corresponding to the beginning and ending times of the interval, respectively. 
These points along  the straight lines are then projected onto the magnetopause interface given by the \citeA{1998JGR...10317691S} model by simply propagating them along the -GSM-X direction. They are then traced along the magnetic field lines based on the \citeA{1996ESASP.389..181T} model to the ionosphere. The series of stars (and nearly overlapping crosses) plotted in Figure~\ref{fig:convmap1} represent these mapped points. They locate around the mapped MMS1 spacecraft position and are also near the region with enhanced poleward flow. There are a total of 20 points confined within the range of the longitudinal extent of the ``opened" flux region spanned by the red dashed curve (see also Figure~\ref{fig:extents}a). Therefore for this event, the axial length of the flux rope at the magnetopause is estimated to be 19 $R_E$.}






\subsection{Event 2: 19 December 2016} 


\begin{figure}
\includegraphics[width=33pc]{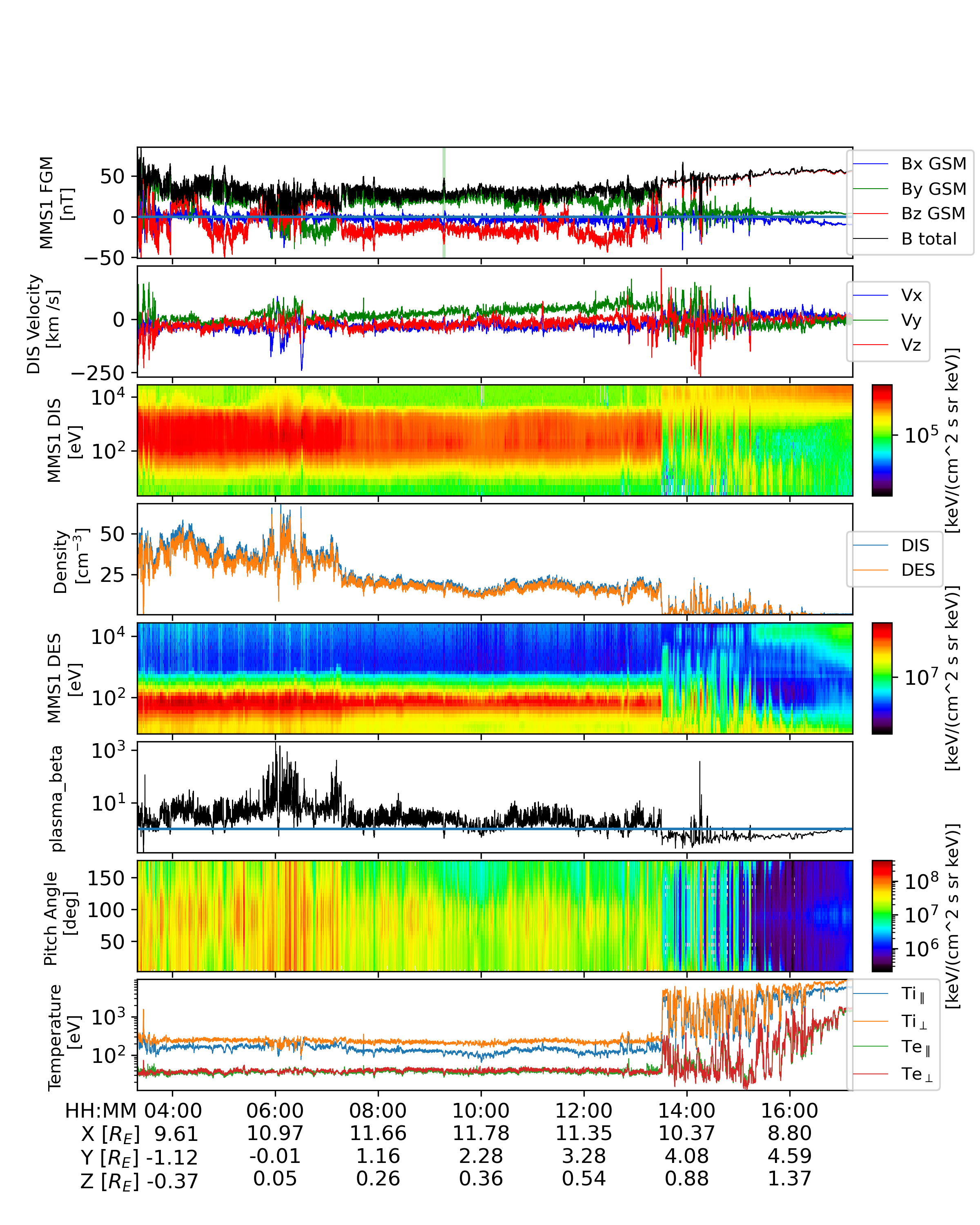}
\caption{Time-series measurements from the MMS1 spacecraft for event 2 on  19 December 2016. Format is the same as Figure~\ref{fig:MMSdata1}. \label{fig:MMSdata2}}
\end{figure}

\begin{figure}
\includegraphics[width=33pc]{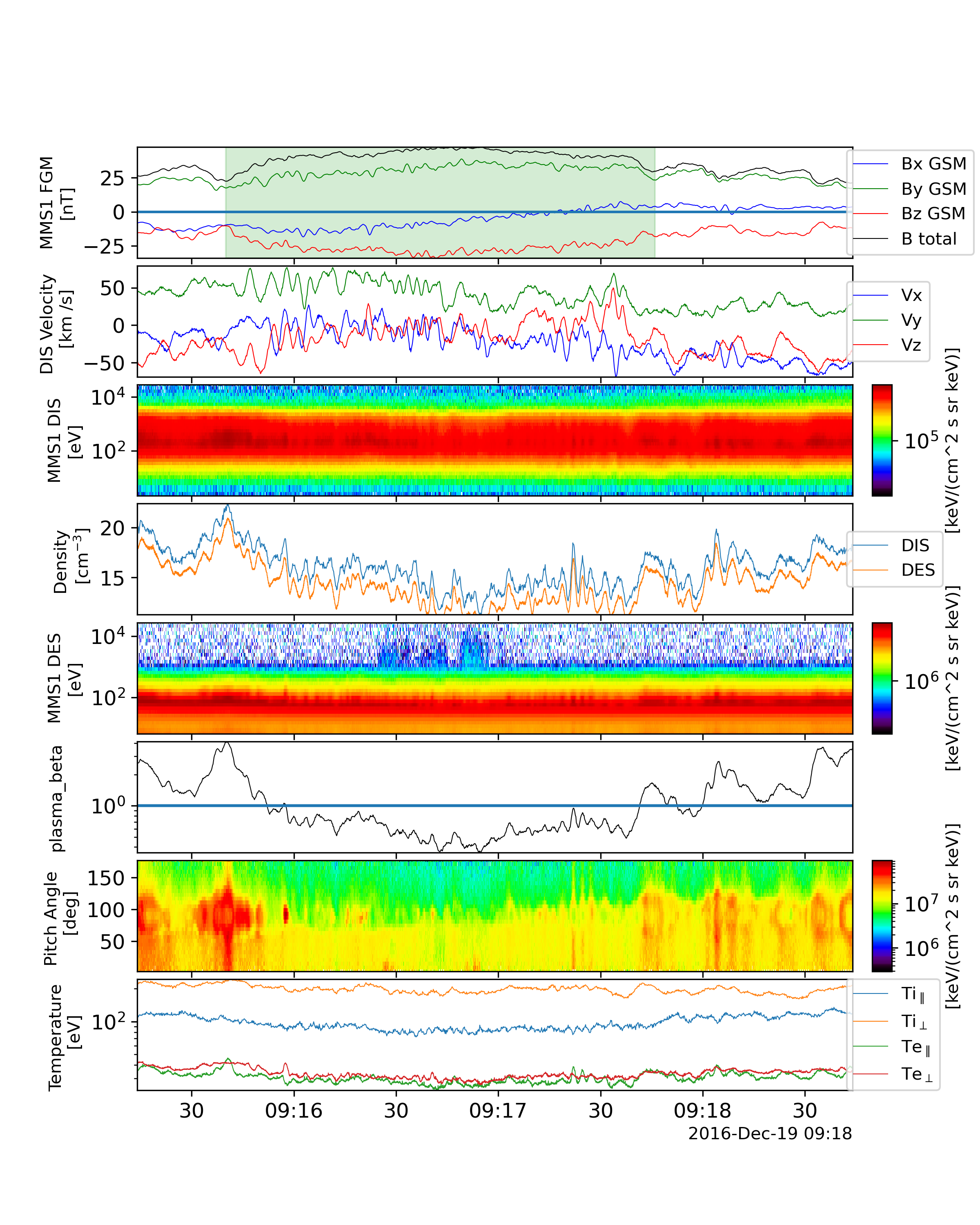}
\caption{Same as Figure~\ref{fig:MMSdata2}, but zoomed in to show the details of the FTE interval, which is marked  by the green shaded area in the top panel. \label{fig:short2}}
\end{figure}

\begin{figure}
\includegraphics[width=.55\textwidth]{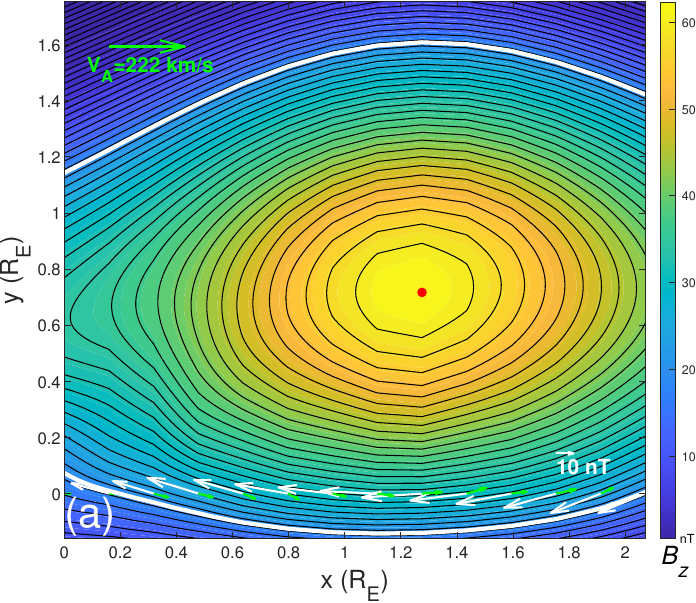}
\includegraphics[width=.44\textwidth]{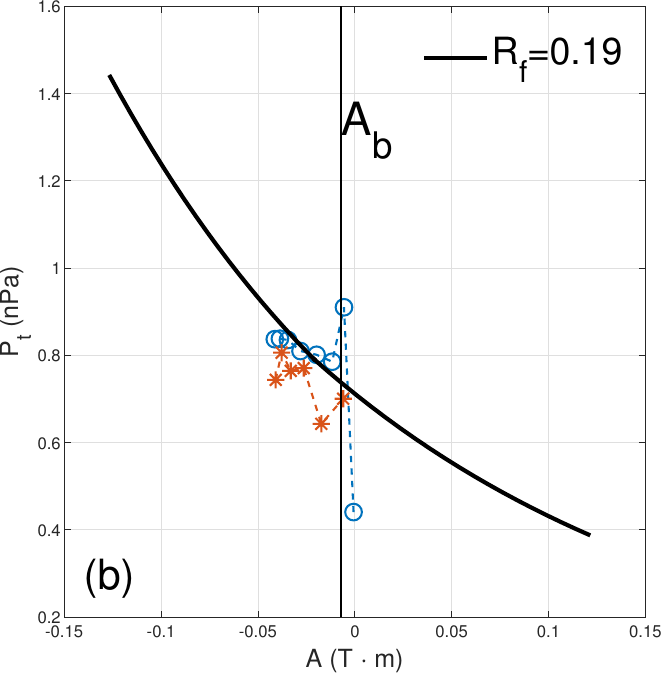}
\caption{The GS reconstruction result for event 2 based on the MMS1 measurements at the magnetopause. Format is the same as Figure~\ref{fig:GSmap1}. The length of the reference vector for the magnetic field is equivalent to 0.25$V_A$.\label{fig:GSmap2}}
\end{figure}

For event 2 on 19 December 2016, the same analysis is carried out. The time series plots are shown in Figures~\ref{fig:MMSdata2} and \ref{fig:short2}. 
The flux rope interval marked in Figure~\ref{fig:short2} shows the magnetic field components with less pronounced rotations in direction, although the field magnitude is elevated. The plasma flow and particle signatures comply with a typical background condition on the magnetosheath side of the magnetopause. The GS reconstruction result for the FTE interval is summarized in Table~\ref{tbl:1} and the cross section map is given in Figure~\ref{fig:GSmap2}a. The results show a flux rope configuration with the $z$ axis mostly pointing southward (i.e., being vertical) in the GSM coordinates. The cross section map consists of closed loops of the contours of the flux function with the increasing $B_z$ value toward the center. The spacecraft path is crossing the edge of the flux rope, corresponding to the insignificant rotation in the field direction. The corresponding $P_t$ versus $A$ curve is shown in Figure~\ref{fig:GSmap2}b, where the fitted functional curve $P_t(A)$ extends significantly  beyond both limits of the range of measurements (i.e., beyond the data points represented by the symbols). The part extrapolated toward the more negative values of $A$ corresponds to the central portion of the flux rope structure enclosed within the white contour shown in Figure~\ref{fig:GSmap2}a. 

{In contrast to event 1,  the ion velocity and spectrum
in Figure~\ref{fig:short2} show no signature of reconnection, the HT velocity in Table~\ref{tbl:1} is too slow to be an encounter with a flux rope part of the FTE, and the
electron perpendicular temperature is higher than the parallel
temperature.  The ePAD lacks clear indication of bi-directional streaming of electrons along the field lines.
All these features suggest that MMS1 detected only the magnetosheath
field lines draping around the FTE flux rope or tube (thus a remote or
grazing encounter). Although such an in-direct encounter for this FTE event is consistent with the GS reconstruction result, given that the flux rope configuration from the GS reconstruction is mostly based on a significant extrapolation of the in-situ data, as described above, the results are thus deemed highly uncertain.}

\begin{figure}
\includegraphics[width=\textwidth]{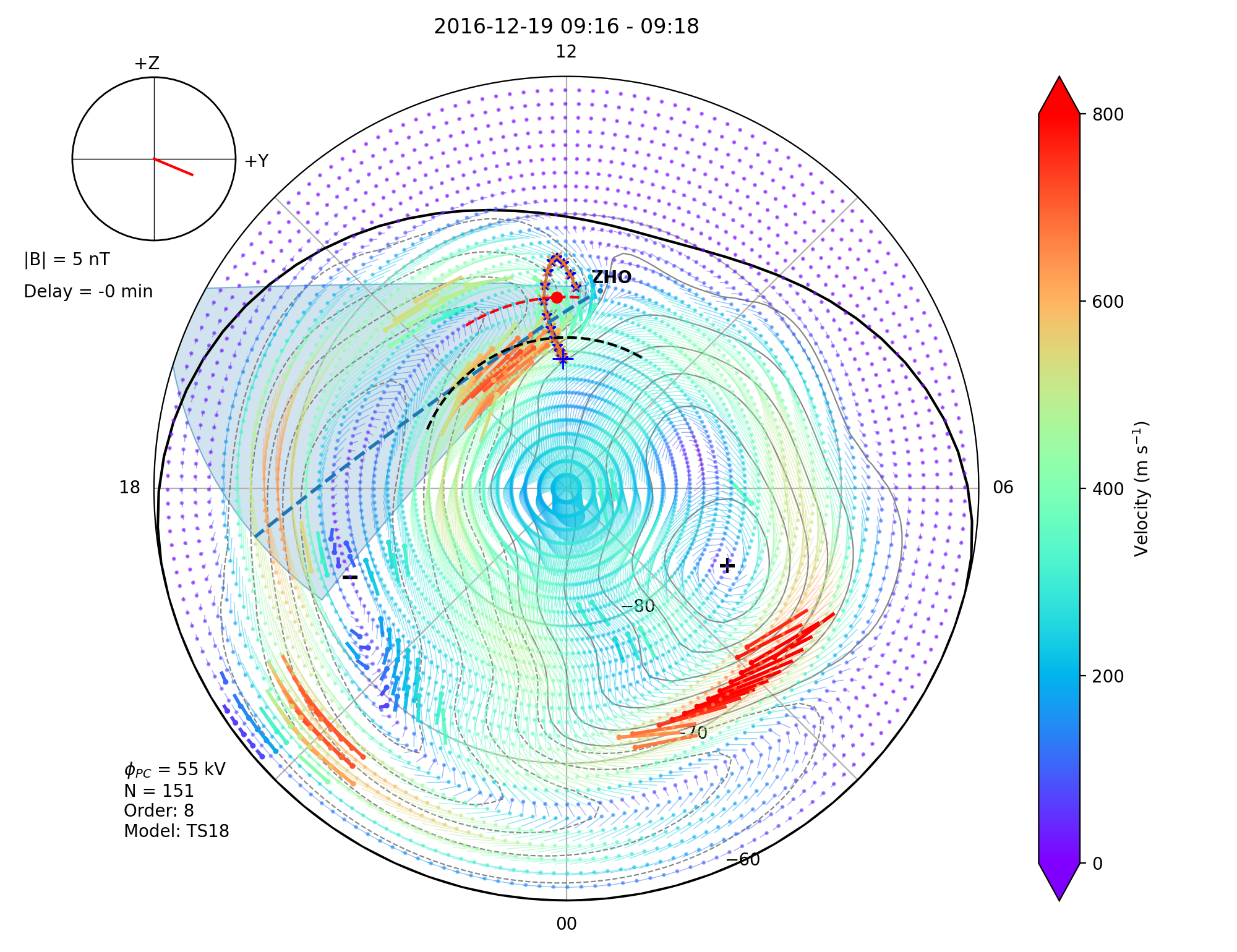}
\caption{Convection map on the ionosphere for event 2. Format is the same as  Figure~\ref{fig:convmap1}.  Points along the axial direction of the reconstructed flux rope projected to the ionosphere are marked as blue stars and red crosses. Each point along the reconstructed flux rope axial direction is 2 $R_{E}$ apart from its neighboring points. The end point marked by the blue plus sign corresponds to the end point along the flux rope axis in the southward direction at the magnetopause.  \label{fig:convmap2}}
\end{figure}

\begin{figure}
\centering
\includegraphics[width=.49\textwidth]{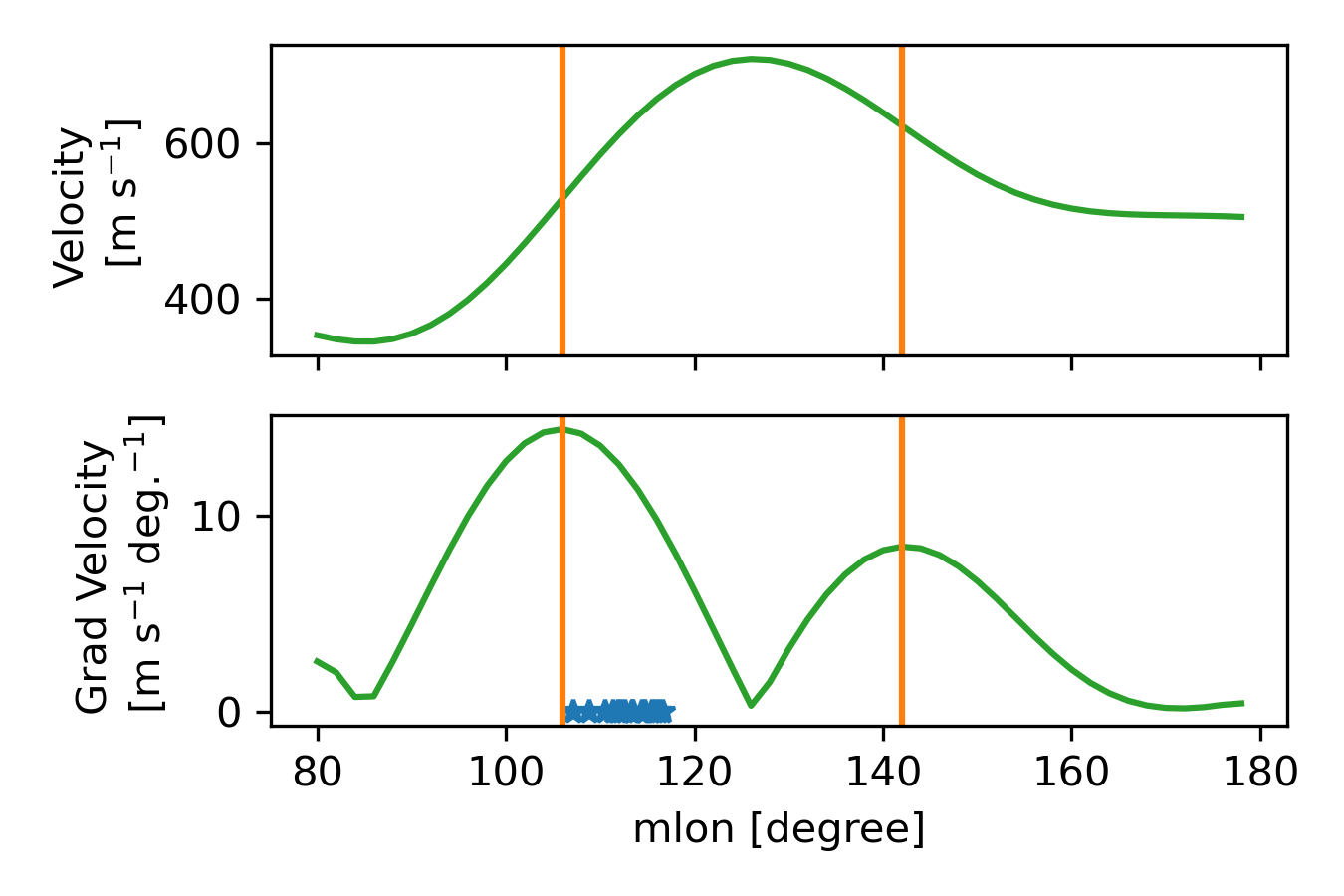}
\includegraphics[width=.5\textwidth]{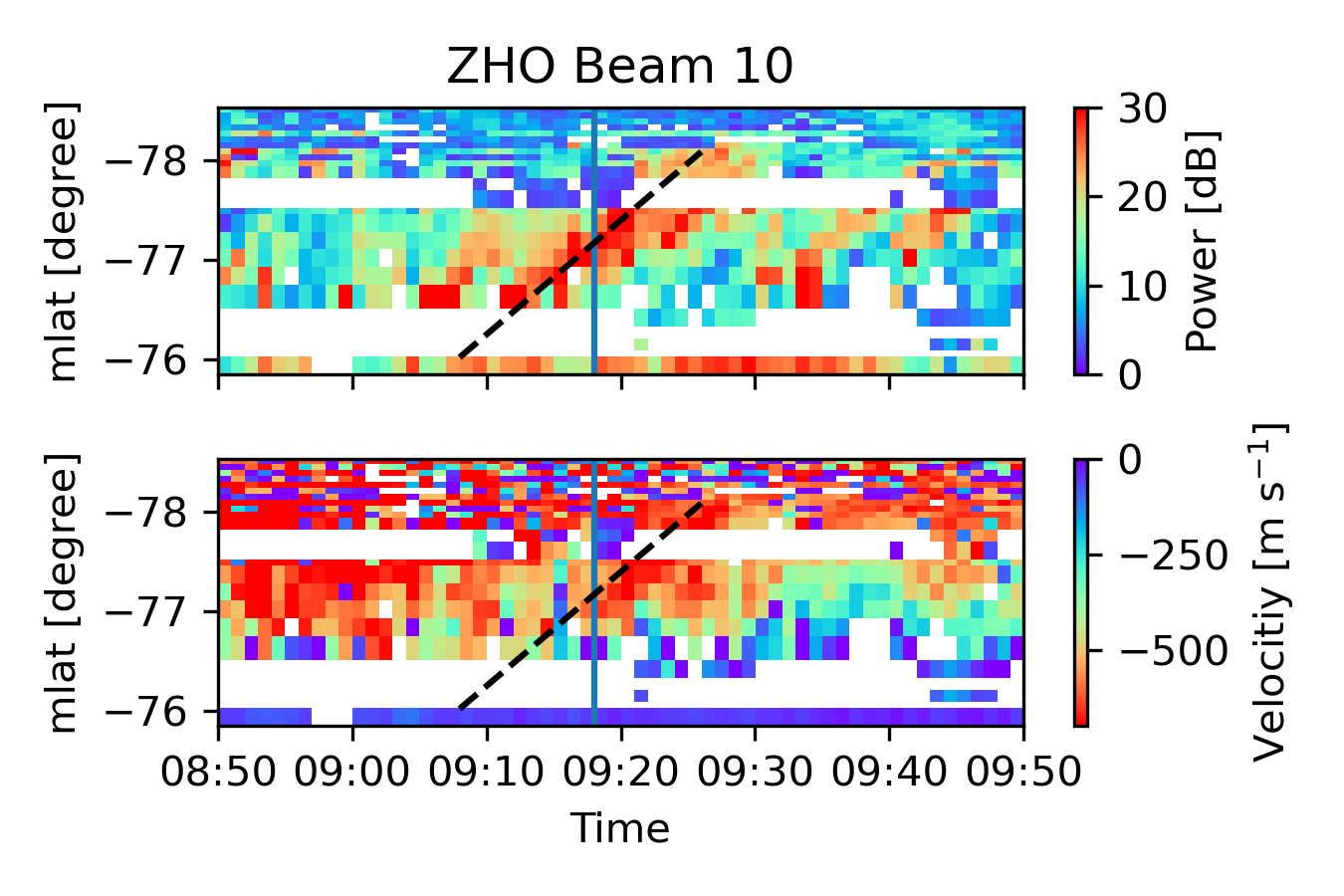}
\hspace{.1\textwidth} (a) \hspace{.4\textwidth} (b)\\
\caption{Analysis of the longitudinal and latitudinal extents of the ``opened" flux region for event 2, based on radar observations from the Zhongshan (ZHO) station. The format is the same as  Figure~\ref{fig:extents}.  }\label{fig:extents2}
\end{figure}

The convection map again from the ZHO station is shown in Figure~\ref{fig:convmap2}, where the poleward enhanced plasma motion is seen near the mapped MMS1 spacecraft position on the ionosphere (12.2 MLT, -76.1$^{\circ}$ MLAT) at the time. The mapped footpoints originating along  the flux rope axis  from the magnetopause to the ionosphere span a relatively narrow range in longitudes, but extend over $\sim 7^\circ$ in latitudes. The analysis based on the radar observations of the back scatter power and the gradient in the convection velocity, shown in Figure~\ref{fig:extents2}, yields a longitudinal extent of 36$^\circ$ and a latitudinal extent of only 2$^\circ$ for the ``opened" area in the polar cap region. Correspondingly, the estimate for the ``opened" flux with uncertainty is 11$\pm$12 MWb, following the same analysis approach as event 1.


\subsection{Summary of GS Reconstruction Results}

\begin{figure}
\centering
\includegraphics[width=1.\textwidth]{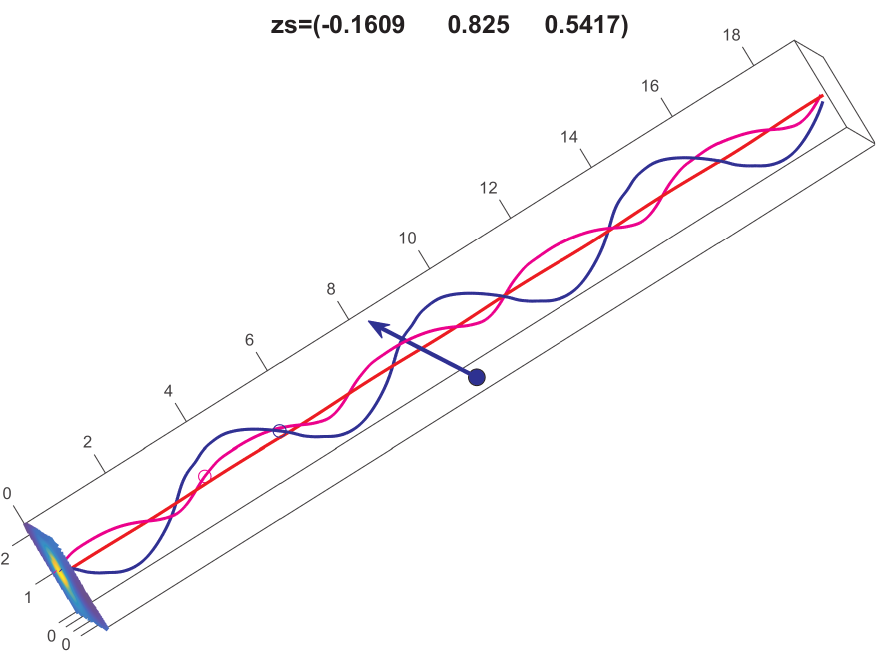}
\caption{A 3D rendering of the field-line configuration for event 1 in a view angle toward Earth, i.e., down  the GSM-X axis. The GSM-Z axis is straight up and the GSM-Y axis is horizontally to the right. The blue dot and arrow denote the location of MMS1 spacecraft at the beginning of the FTE interval and the direction of $-\mathbf{V}_{HT}$, respectively.   The tickmark labels are in $R_E$. Three field lines are drawn in red, magenta and blue colors within the cylindrical volume of axial length $L=19 R_E$ based on our analysis results for event 1. The cross section map is shown on the bottom plane where the field lines are rooted and the optimal $z$ axis orientation in the GSM coordinates is denoted on top.} \label{fig:3Dview}
\end{figure}

\begin{table}[htb]
 \caption{Analysis results based on the GS reconstruction of the FTE flux rope interval at the magnetopause and the corresponding radar observations in the ionosphere for event 1$^a$.}\label{tbl:event1}
 {\centering
 \begin{tabular}{l l l l l l l}
 \hline
 $\Phi_z$ [MWb]& $\phi_p$ [T$\cdot$m] & $L$ [$R_E$] &$\Phi_p$ [MWb]& $\Phi_R$ [MWb]&$\tilde\phi_p$ [T$\cdot$m] &$\tilde\Phi_z$ [MWb]\\
 \hline
 3.4 & 0.107 & 19 & 13 & 28$\pm$16 &0.0268 - 0.0621 & 1.05 - 3.59\\
 \hline
\end{tabular}}
{\footnotesize{$^{a}$The last two columns are the ranges of  fluxes cited from \citeA{2006AnGeo..24..603H} for 5 FTEs at the magnetopause by applying the optimal GS reconstruction method to the Cluster spacecraft data. }}
\end{table}
We summarize the GS reconstruction results, mainly the magnetic flux estimates in Table~\ref{tbl:event1} for event 1 only, because those results are judged to be reliable based on the analysis results presented in Section~\ref{sec:event1analysis}. 
The ranges of the unit polodial flux and the axial flux for five FTE intervals examined by \citeA{2006AnGeo..24..603H} are also shown for comparison, where the maximum values for both fluxes are from one FTE interval. The axial flux of event 1 is within the range of those estimates, although closer to the upper limit. The unit poloidal flux is about 50\% larger than the upper limit of the range of the corresponding estimates from  \citeA{2006AnGeo..24..603H}.

The largest uncertainty in the estimate of the total poloidal flux $\Phi_p$ is generally believed to lie in the uncertainty of the axial length, $L$, of a cylindrical flux rope model \cite{2014ApJH,2015JGRAH}. In this analysis, we lack a feasible means to provide an estimate of the uncertainty associated with $L$. If we adopt the same assumption as we made for the similar analysis of the interplanetary magnetic flux ropes \cite{2014ApJH,2015JGRAH}, the uncertainty in $\Phi_p$ could amount to 100\% toward the estimate of the upper limit of the total poloidal flux.

To further put our GS reconstruction result in the context of a better characterization of the magnetic field configuration of an FTE flux rope from a quasi-3D point of view, we show in Figure~\ref{fig:3Dview} a rendering of the 3D field line plot for event 1. It is the solution to the GS equation within the solution domain of a cylinder (or a cuboid)  with the axial length $L=19 R_E$. The cylinder is oriented along the $z$ axis direction as viewed toward the Earth with the dawn-dusk direction pointing horizontally to the right. The MMS1 spacecraft is traversing the structure along the blue arrow at the time with the velocity, $-\mathbf{V}_{HT}$, given in Table~\ref{tbl:1}. In other words, the structure is moving with the velocity  $\mathbf{V}_{HT}$ relative to the spacecraft in the opposite direction of the blue arrow. Three selected field lines are drawn. The straight line in red originates from the red dot in Figure~\ref{fig:GSmap1}a, where $B_z$ reaches the maximum. The magenta and blue lines are spiraling along the $z$ axis around the central line with varying degrees of twist. On average, the unit field line twist can be estimated by taking the ratio between $\phi_p$ and $\Phi_z$ \cite{2014ApJH}, which yields about 0.2 turns/$R_E$ for event 1. Therefore for the flux rope configuration shown in Figure~\ref{fig:3Dview} with $L=19 R_E$, the average total number of twist or turns of the field lines for the FTE flux rope is approximately 4. 

\section{Interpretation for the formation of magnetic flux ropes at the magnetopause}\label{sec:interp}
Based on these analysis results, we would like to describe in detail our view on the FTE flux rope formation as this is the main motivation for this study.
The basic process to be proposed for the formation of magnetic flux ropes at the Earth's magnetopause, i.e., in the form of FTEs, is largely based on the well-known scenario of flux rope formation on the Sun through magnetic reconnection as manifested in solar flares \cite{Longcope2007b,Qiu2007,2014ApJH}. An analogy between the topological change of the underlying magnetic field lines during a solar flare and that during the FTE formation can be made because the magnetic reconnection is responsible for these changes in both cases. As a result, a  common magnetic flux rope structure  is formed in both cases, albeit it is drastically different in size and strength (or magnetic flux content). In addition, these changes, as reflected by the reconnected magnetic field line footpoints motion, can be both characterized by remote-sensing observations. For solar flares, they are generally represented by the flare ribbon brightenings primarily observed on the chromosphere where the reconnected field lines map to and exhibit enhanced brightenning in patches during a flare.  For FTEs, the reconnected field line footpoints may map to the ionosphere, causing enhanced convection flows that can be measured by, e.g., the SuperDARN radar network in polar regions. We provide, in this study, an interpretation of the FTE formation at the Earth's magnetopause based on such an analogy between the magnetic field topologies of the two processes.

\begin{figure}
\includegraphics[width=\textwidth]{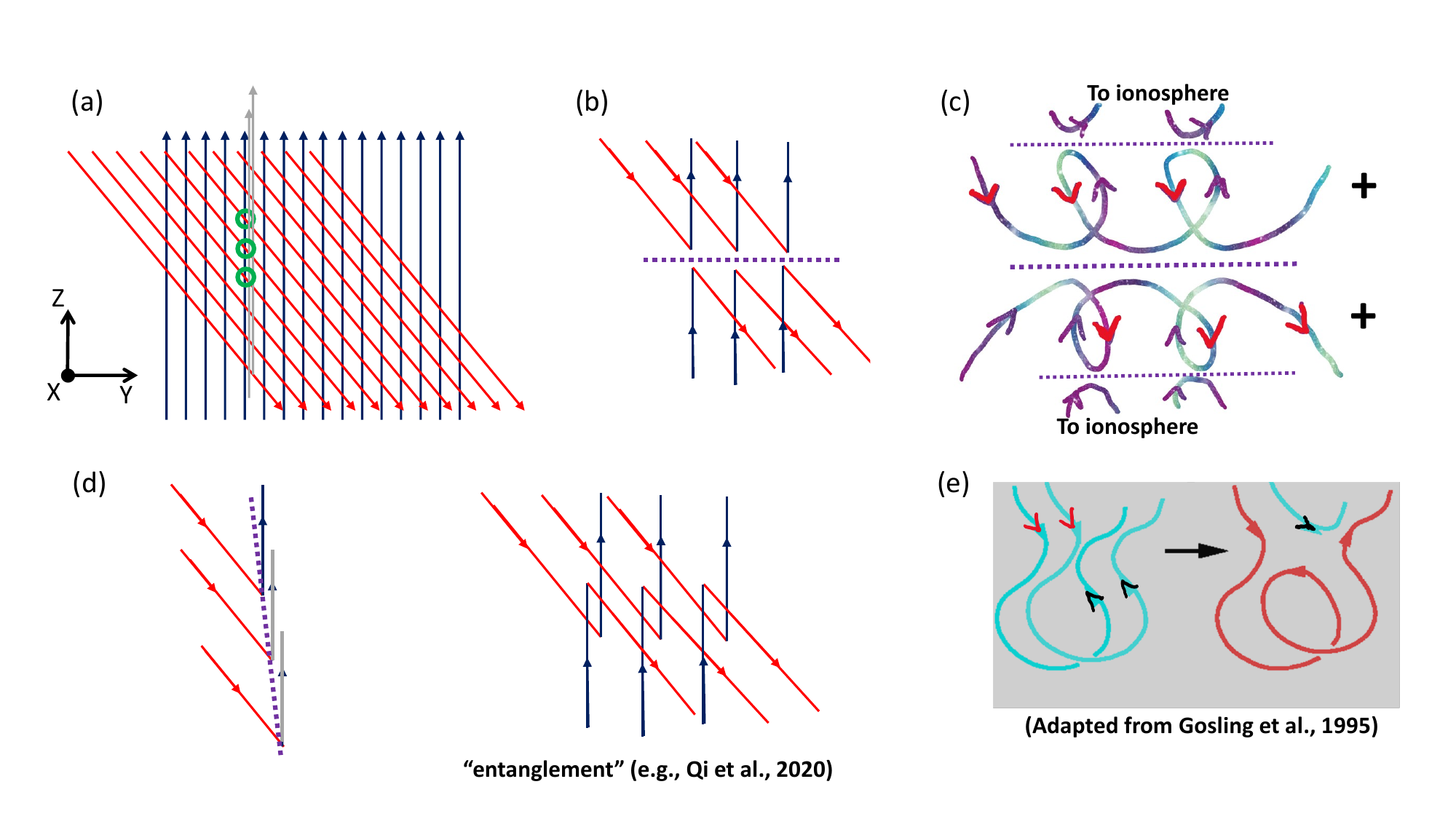}
\caption{Schematics for the formation of magnetic flux ropes at the Earth's magnetopause. (a) The magnetic field lines in the magnetosheath (red lines) and the magnetosphere (black lines) as viewed toward Earth (along -X) at the magnetopause for a general IMF condition of $B_Z<0$ and $B_Y>0$ where the X, Y, and Z donote the unit directional vectors of the GSM coordinates. This view can also be considered equivalent to an LMN coordinate system, e.g., with L$\equiv$Z, M$\equiv$-Y, and N$\equiv$X. (b) One scenario of magnetic  reconnection between the field lines in the magnetosheath and in  the magnetosphere, forming a primary X line as denoted by the thick dotted line. (c) The subsequent reconnection between adjacent field lines along the secondary X lines (thin dotted lines) forming twisted magnetic flux rope structures which are right-handed as denoted by the ``+" sign. The other sets of reconnected field lines as represented by shorter loops may connect to the ionosphere. (d) One distinctive scenario of  an approximately north-south oriented primary X line, and an alternative scenario of flux ``entanglement" \cite{2020GL090314entanglement} with the red line segments always crossing above the black line segments in this view (see text for details). (e) The detailed change of connectivity for the reconnection between one pair of adjacent field lines as originally depicted by \citeA{1995GeoRL..22..869G}. The viewpoint is the same for all panels except for (e), and the red arrowheads are consistently attached to the parts of field lines corresponding to the original red field lines in (a) and (b).  } \label{fig:MFR}
\end{figure}
Figure~\ref{fig:MFR} shows schematic illustrations of the possible scenario of FTE formation through a sequential reconnection process at the magnetopause under a southward IMF (or magnetosheath field) condition with $B_Y>0$. A similar set can be generated for the similar IMF condition but with $B_Y<0$. Such a process, especially as illustrated in panel (c), has a direct analogy to the formation of a flux rope with quasi-3D geometry in solar flares \cite{Longcope2007b,Qiu2007,2014ApJH}. Figure~\ref{fig:MFR}a shows the relative orientations of the fields in the magnetosheath (red lines) and the magnetosphere (black lines). In general magnetic reconnection may proceed between any pair of the red and black field lines, forming an arbitrary primary X line. Figure~\ref{fig:MFR}b shows  a typical case with an approximately dawn-dusk (horizontally) oriented primary X line along which two sets of reconnected field lines are aligned in a crisscross pattern approximately in the same horizontal direction. Subsequently, as illustrated in (c),  additional reconnection may ensue for either or both sets of field lines separated by the primary X line, especially in a sequence that happens between adjacent field lines and proceeds either from dawn toward dusk direction or vice versa. 

For example, for the set of field lines north of the primary X line illustrated in Figure~\ref{fig:MFR}b, assume that reconnection proceeds from dawn to dusk (left to right), the black end of the first field line may approach the red end of the adjacent field line and reconnect, forming a twisted field line. Such a process is further shown in Figure~\ref{fig:MFR}e \cite{1995GeoRL..22..869G}, which has long been recognized as a building block for the formation of a magnetic flux rope configuration with field lines of multiple turns, as depicted in (c), as a result of such a sequential reconnection process between adjacent field lines. If a ``guide field" can be assumed to be determined by the $B_Y$ component, dictating the formation of an axial field near the center of such a flux rope configuration, the handedness (the sign of magnetic helility, or chirality) of the magnetic flux rope topology can be inferred as right-handed (being positive in chirality, ``+") for either set of field lines separated by the primary X line in Figure~\ref{fig:MFR}c and for either direction of the reconnection sequence. It can also be shown that for the other condition $B_Y<0$, a left-handed magnetic flux rope (negative chirality, ``-")  may form in a similar manner with an approximately horizontal orientation. These findings about handedness rules of FTE flux ropes are consistent with the recent studies of the sign of helicity of FTEs \cite{2020GL091257K,2022JA030686D}. They concluded that ``Right-handed (left-handed) FTE flux ropes are mostly preceded by positive (negative) interplanetary magnetic field (IMF) $B_Y$" as one of their key points \cite{2020GL091257K}. Furthermore, it was pointed out by \citeA{2022JA030686D} that a weaker $B_Y$ component corresponding to the lack of a ``guide field" may lead to greater uncertainty in the aforementioned chirality rule, although those authors attributed such uncertainty to microscopic (Hall) effect \cite<see, also,>[]{Eriksson2020JA027919}. However we would argue that the complexity in a more general 3D magnetic field topology may disrupt the chirality rule when the field configuration becomes complex and deviates from a 2D geometry.

{Topologically, Figure~\ref{fig:MFR}d illustrates such a special case or an exception to the aforementioned chirality rule, corresponding to a case with a north-south (vertically) oriented primary X line. For this case, it is not clear how a ``guide field" contributes to the formation of the axial field of a 2D magnetic flux rope. Such a configuration shown to the left may be formed through consecutive reconnection at the locations marked by the green circles in (a), where the reconnection proceeds between the red field lines and the corresponding black lines at the sites aligned vertically with the black field lines replenished from the magnetosphere as indicated by the additional gray lines. Whereas perhaps a more common scenario, given to the right in (d), is the flux ``entanglement" \cite{2020GL090314entanglement}, or inter-laced/linked field lines \cite{2020GeoRL..4786726F}, due to the fact of the prevalent intercepting X points originating from (a). Both configurations impose significant difficulty for the GS reconstruction method \cite<see, e.g.,>[]{2007JA012492HH}.
Additionally, in a more general 3D topology that goes beyond the current most commonly invoked 2D framework, the uncertainty in the handedness is probably intrinsic to the complexity in the field topology, leading to further complications in characterizing the magnetic flux contents as well. }

In summary, as a consequence of such a reconnection sequence in a quasi-3D geometry, in the case of flares, the amount of magnetic flux enclosed by regions swept by the flare ribbons (so-called the reconnection flux) usually matches the poloidal flux of thus formed magnetic flux rope \cite{Qiu2007,2014ApJH}. In direct analogy, the same applies for the connection between the FTE flux rope and the region in the ionosphere where the reconnected field lines sweep through. 
Specifically, from Figure~\ref{fig:MFR}c and e, it can be understood that one unit of magnetic flux injected into the coiled loop structures adding one turn of the ensuing flux rope is equal to the amount closed down into the reconnected short loop with one end tracing to the ionosphere. Therefore, the amount of poloidal flux of thus formed FTE flux rope corresponds to the amount of flux ``opened" (i.e., the reconnection flux $\Phi_R$) in the corresponding polar cap region where the reconnected field line footpoints are rooted. 

\section{Conclusions and Discussion}\label{sec:dis}
In conclusion, we have presented two event studies of the FTE flux ropes at the Earth's magnetopause based on the MMS1 in-situ measurements and the corresponding mapped field-line footpoint motion in the high-latitude ionosphere in the Southern Hemisphere based on the simultaneous SuperDARN radar observations. The GS reconstruction method is applied to the in-situ measurements of FTE flux ropes to derive the magnetic field configuration in a cylindrical geometry, which yields the quantitative characterizations of the magnetic flux contents of the flux rope structure, in terms of the toroidal (axial) and poloidal magnetic fluxes. In turn, the corresponding reconnection flux in an area is estimated by examining the correlated enhanced plasma convection pattern mapped along the magnetospheric field lines to the polar cap region, following the approach of \citeA{2017JGRA..12212310F}. The area ``opened" through the magnetic reconnection at the dayside magnetopause, forming the FTE, and mapped to the ionosphere is estimated by calculating the longitudinal and latitudinal extents based on the SuperDARN observations of nearly concurrent enhancement of poleward plasma convection motion during the FTE interval.

We find that for event 1, the FTE flux rope configuration is well reconstructed with the spacecraft path cutting across the center of a helical magnetic field structure. It possesses right-handed chirality and is oriented largely in the dawn-dusk direction at the magnetopause. The flux rope length is estimated to be about 19 $R_E$. The GS reconstruction results yield the corresponding magnetic flux contents as listed in Table~\ref{tbl:event1}. The poloidal flux of the FTE flux rope, $\Phi_p=$13 MWb, falls within the range of the estimated reconnection flux, $\Phi_R=$28$\pm$16 MWb, from the SuperDARN radar observations, but the toroidal (axial) flux is significantly lower. Considering the possible uncertainty in the estimation of the flux rope length, $L$, and its variability (e.g., in \citeA{2017JGRA..12212310F}, such a length for one event was estimated to be as large as 38 $R_E$), the agreement between the poloidal flux of the FTE flux rope and the reconnection flux for this event is likely supported by these analysis results. 

{For event 2, as indicated in Table~\ref{tbl:1}, the FTE flux rope possesses an axial orientation that is in the North-South direction and left-handed chirality, despite the fact that the spacecraft path is near the edge of the flux rope cross section, as shown in Figure~\ref{fig:GSmap2}a. Both the axial and the poloidal fluxes are comparable to the values for event 1. However, they are considered less reliable due to the fact that the flux rope configuration is derived mainly by extrapolations to the in-situ spacecraft data. There also exists significant uncertainty in the estimate of the reconnection flux from radar observations. Additional source of uncertainty is associated with the estimate of the axial length of the flux rope. Since the flux rope is mostly oriented vertically at the magnetopause from a viewpoint toward the Earth in the GSM coordinates, the mapped field line footpoints from the flux rope to the ionosphere, as shown in Figure~\ref{fig:convmap2}, tend to congregate around the same longitude, resulting in a larger uncertainty in the flux rope length, $L$. Therefore, all these aforementioned uncertainties for event 2 prohibit a quantitative comparison among the various flux estimates.}

Motivated by the conceptual analogy to solar flares, we offer an interpretation of the formation of the FTE flux rope through the magnetic reconnection at the dayside magnetopause, as presented in Section~\ref{sec:interp}. Figure~\ref{fig:MFR} illustrates a scenario for the IMF $B_Y>0$, which is consistent with the results for events 1 summarized above. In particular, it provides a more detailed explanation for the chirality rule of FTE flux ropes  \cite<e.g.,>[]{2022JA030686D,2020GL091257K} based on the topology change of FTE flux ropes through magnetic reconnection. The distinction of this scenario from the others \cite<e.g.,>[and others]{1985GeoRL..12..105L} is perhaps the emphasis on the intermediate process (i.e., that generating the shorter loops marked by ``To ionosphere" in Figure~\ref{fig:MFR}c), corresponding to the reconnection sequence between adjacent reconnecting field lines from one end to the other. The entire sequence may generate the corresponding signatures in the ionosphere, not just at the two ends. 
{Therefore we conclude that the flux rope formation at the magnetopause may proceed in a quasi-3D manner via a sequential magnetic reconnection process between adjacent field-line loops. Such a sequence may dictate the topological properties of the thus formed magnetic flux rope, governed by the IMF condition and other spatial features, such as the orientations of the multiple X-lines (see Figure~\ref{fig:MFR}). The results from event 1, especially in terms of the agreement between the poloidal flux and the reconnection flux and the correct handedness, support this conclusion, while for event 2, it is much uncertain. It indicates the importance of detailed investigation of magnetic field topology into two or three dimensions that has to go beyond a relatively simple time-series analysis often limited to one spatial dimension.}

To further  elucidate these points, 
we will extend the current study which is limited by the small number of event studies. A survey of additional FTE events with conjugate signatures in the ionosphere by employing the approaches described here (or the ones with refined analysis to reduce uncertainties) can be carried out in the future. 
In addition, it has been increasingly realized that in a more general 3D topology, the reconnection sequence may indicate a correlation between the axial flux of the flux rope and the reconnection flux  \cite<see, e.g.,>[]{he2022quantitative,hu2022validation}. Therefore it is worth investigating further the correlation between the reconnection flux and the flux encompassed in FTE events, as this study and other previous studies have attempted to do, through multiple observational and theoretical approaches. 

\section*{Data Availability Statement}
SuperDARN data can be found at \url{https://www.frdr-dfdr.ca/repo/collection/superdarn}. SuperDARN data has been processed using the Radar Software Toolkit developed by the SuperDARN Data Analysis Working Group \cite{superdarn_data_analysis_working_group_2022_6473603} and visualized by the pyDARN package  developed by the SuperDARN Data Visualization Working Group \cite{superdarn_data_visualization_working_gro_2023_7767590,Shifspas.2022.1022690}. The MMS spacecraft data are accessed via the MMS Science Data Center (\url{https://lasp.colorado.edu/mms/sdc/public/}). A Python package, pyGS, developed by Dr.~Yu Chen for performing the GS reconstruction, is publicly available at \url{https://github.com/PyGSDR/PyGS/}.

\acknowledgments
We acknowledge the use of SuperDARN data. The Zhongshan (ZHO) SuperDARN radar is maintained and operated by the Polar Research Institute of China with partial support from the Chinese Meridian Project.  SuperDARN is a network of radars funded by national scientific funding agencies of Australia, Canada, China, France, Italy, Japan, Norway, South Africa, the United Kingdom, and the United States of America. SW, YZ and QH acknowledge partial support of NASA grant 80NSSC21K0003. XS is supported by NSF grants AGS-1935110 and AGS-2025570, and NASA grant 80NSSC21K1677.  The work by H.H. was supported by JSPS Grant-in-aid for Scientific
Research KAKENHI 21K03504.

\bibliography{ref_master3}

\begin{thebibliography}{}

\bibitem [\protect \citeauthoryear {%
Burrell%
\ \protect \BOthers {.}}{%
Burrell%
\ \protect \BOthers {.}}{%
{\protect \APACyear {2022}}%
}]{%
superdarn_data_analysis_working_group_2022_6473603}
\APACinsertmetastar {%
superdarn_data_analysis_working_group_2022_6473603}%
\begin{APACrefauthors}%
Burrell, A.%
, Thomas, E.%
, Schmidt, M.%
, Bland, E.%
, Coco, I.%
, Ponomarenko, P.%
\BDBL {}Walach, M\BHBI T.%
\end{APACrefauthors}%
\unskip\
\newblock
\APACrefYearMonthDay{2022}{{\APACmonth{04}}}{}.
\newblock
\APACrefbtitle {SuperDARN Radar Software Toolkit (RST) 4.7.} {Superdarn radar
  software toolkit (rst) 4.7.}
\newblock
\APACaddressPublisher{}{Zenodo}.
\newblock
\begin{APACrefURL} \url{https://doi.org/10.5281/zenodo.6473603}
  \end{APACrefURL}
\newblock
\begin{APACrefDOI} \doi{10.5281/zenodo.6473603} \end{APACrefDOI}
\PrintBackRefs{\CurrentBib}

\bibitem [\protect \citeauthoryear {%
{Chen}%
}{%
{Chen}%
}{%
{\protect \APACyear {2011}}%
}]{%
2011LRSP....8....1C}
\APACinsertmetastar {%
2011LRSP....8....1C}%
\begin{APACrefauthors}%
{Chen}, P\BPBI F.%
\end{APACrefauthors}%
\unskip\
\newblock
\APACrefYearMonthDay{2011}{{\APACmonth{04}}}{}.
\newblock
{\BBOQ}\APACrefatitle {{Coronal Mass Ejections: Models and Their Observational
  Basis}} {{Coronal Mass Ejections: Models and Their Observational
  Basis}}.{\BBCQ}
\newblock
\APACjournalVolNumPages{Living Reviews in Solar Physics}{8}{1}{1}.
\newblock
\begin{APACrefDOI} \doi{10.12942/lrsp-2011-1} \end{APACrefDOI}
\PrintBackRefs{\CurrentBib}

\bibitem [\protect \citeauthoryear {%
{Chisham}%
\ \protect \BOthers {.}}{%
{Chisham}%
\ \protect \BOthers {.}}{%
{\protect \APACyear {2007}}%
}]{%
2007SGeo...28...33C}
\APACinsertmetastar {%
2007SGeo...28...33C}%
\begin{APACrefauthors}%
{Chisham}, G.%
, {Lester}, M.%
, {Milan}, S\BPBI E.%
, {Freeman}, M\BPBI P.%
, {Bristow}, W\BPBI A.%
, {Grocott}, A.%
\BDBL {}{Walker}, A\BPBI D\BPBI M.%
\end{APACrefauthors}%
\unskip\
\newblock
\APACrefYearMonthDay{2007}{{\APACmonth{01}}}{}.
\newblock
{\BBOQ}\APACrefatitle {{A decade of the Super Dual Auroral Radar Network
  (SuperDARN): scientific achievements, new techniques and future directions}}
  {{A decade of the Super Dual Auroral Radar Network (SuperDARN): scientific
  achievements, new techniques and future directions}}.{\BBCQ}
\newblock
\APACjournalVolNumPages{Surveys in Geophysics}{28}{1}{33-109}.
\newblock
\begin{APACrefDOI} \doi{10.1007/s10712-007-9017-8} \end{APACrefDOI}
\PrintBackRefs{\CurrentBib}

\bibitem [\protect \citeauthoryear {%
Dahani%
\ \protect \BOthers {.}}{%
Dahani%
\ \protect \BOthers {.}}{%
{\protect \APACyear {2022}}%
}]{%
2022JA030686D}
\APACinsertmetastar {%
2022JA030686D}%
\begin{APACrefauthors}%
Dahani, S.%
, Kieokaew, R.%
, Génot, V.%
, Lavraud, B.%
, Chen, Y.%
, Michotte~de Welle, B.%
\BDBL {}Burch, J.%
\end{APACrefauthors}%
\unskip\
\newblock
\APACrefYearMonthDay{2022}{}{}.
\newblock
{\BBOQ}\APACrefatitle {The Helicity Sign of Flux Transfer Event Flux Ropes and
  Its Relationship to the Guide Field and Hall Physics in Magnetic Reconnection
  at the Magnetopause} {The helicity sign of flux transfer event flux ropes and
  its relationship to the guide field and hall physics in magnetic reconnection
  at the magnetopause}.{\BBCQ}
\newblock
\APACjournalVolNumPages{Journal of Geophysical Research: Space
  Physics}{127}{11}{e2022JA030686}.
\newblock
\begin{APACrefURL}
  \url{https://agupubs.onlinelibrary.wiley.com/doi/abs/10.1029/2022JA030686}
  \end{APACrefURL}
\newblock
\APACrefnote{e2022JA030686 2022JA030686}
\newblock
\begin{APACrefDOI} \doi{https://doi.org/10.1029/2022JA030686} \end{APACrefDOI}
\PrintBackRefs{\CurrentBib}

\bibitem [\protect \citeauthoryear {%
{Elphic}%
}{%
{Elphic}%
}{%
{\protect \APACyear {1990}}%
}]{%
1990GMS....58..455E}
\APACinsertmetastar {%
1990GMS....58..455E}%
\begin{APACrefauthors}%
{Elphic}, R\BPBI C.%
\end{APACrefauthors}%
\unskip\
\newblock
\APACrefYearMonthDay{1990}{{\APACmonth{01}}}{}.
\newblock
{\BBOQ}\APACrefatitle {{Observations of flux transfer events - Are FTEs flux
  ropes, islands, or surface waves?}} {{Observations of flux transfer events -
  Are FTEs flux ropes, islands, or surface waves?}}{\BBCQ}
\newblock
\APACjournalVolNumPages{Geophysical Monograph Series}{}{}{455-471}.
\PrintBackRefs{\CurrentBib}

\bibitem [\protect \citeauthoryear {%
{Elphic}%
, {Lockwood}%
, {Cowley}%
\BCBL {}\ \BBA {} {Sandholt}%
}{%
{Elphic}%
\ \protect \BOthers {.}}{%
{\protect \APACyear {1990}}%
}]{%
1990GeoRL..17.2241E}
\APACinsertmetastar {%
1990GeoRL..17.2241E}%
\begin{APACrefauthors}%
{Elphic}, R\BPBI C.%
, {Lockwood}, M.%
, {Cowley}, S\BPBI W\BPBI H.%
\BCBL {}\ \BBA {} {Sandholt}, P\BPBI E.%
\end{APACrefauthors}%
\unskip\
\newblock
\APACrefYearMonthDay{1990}{{\APACmonth{11}}}{}.
\newblock
{\BBOQ}\APACrefatitle {{Flux transfer events at the magnetopause and in the
  ionosphere}} {{Flux transfer events at the magnetopause and in the
  ionosphere}}.{\BBCQ}
\newblock
\APACjournalVolNumPages{Geophysical Research Letters}{17}{12}{2241-2244}.
\newblock
\begin{APACrefDOI} \doi{10.1029/GL017i012p02241} \end{APACrefDOI}
\PrintBackRefs{\CurrentBib}

\bibitem [\protect \citeauthoryear {%
Eriksson%
, Souza%
, Cassak%
\BCBL {}\ \BBA {} Hoilijoki%
}{%
Eriksson%
\ \protect \BOthers {.}}{%
{\protect \APACyear {2020}}%
}]{%
Eriksson2020JA027919}
\APACinsertmetastar {%
Eriksson2020JA027919}%
\begin{APACrefauthors}%
Eriksson, S.%
, Souza, V\BPBI M.%
, Cassak, P\BPBI A.%
\BCBL {}\ \BBA {} Hoilijoki, S.%
\end{APACrefauthors}%
\unskip\
\newblock
\APACrefYearMonthDay{2020}{}{}.
\newblock
{\BBOQ}\APACrefatitle {Nascent Flux Rope Observations at Earth's Dayside
  Magnetopause} {Nascent flux rope observations at earth's dayside
  magnetopause}.{\BBCQ}
\newblock
\APACjournalVolNumPages{Journal of Geophysical Research: Space
  Physics}{125}{10}{e2020JA027919}.
\newblock
\begin{APACrefURL}
  \url{https://agupubs.onlinelibrary.wiley.com/doi/abs/10.1029/2020JA027919}
  \end{APACrefURL}
\newblock
\APACrefnote{e2020JA027919 2020JA027919}
\newblock
\begin{APACrefDOI} \doi{https://doi.org/10.1029/2020JA027919} \end{APACrefDOI}
\PrintBackRefs{\CurrentBib}

\bibitem [\protect \citeauthoryear {%
{Fargette}%
\ \protect \BOthers {.}}{%
{Fargette}%
\ \protect \BOthers {.}}{%
{\protect \APACyear {2020}}%
}]{%
2020GeoRL..4786726F}
\APACinsertmetastar {%
2020GeoRL..4786726F}%
\begin{APACrefauthors}%
{Fargette}, N.%
, {Lavraud}, B.%
, {{\O}ieroset}, M.%
, {Phan}, T\BPBI D.%
, {Toledo-Redondo}, S.%
, {Kieokaew}, R.%
\BDBL {}{Smith}, S\BPBI E.%
\end{APACrefauthors}%
\unskip\
\newblock
\APACrefYearMonthDay{2020}{{\APACmonth{03}}}{}.
\newblock
{\BBOQ}\APACrefatitle {{On the Ubiquity of Magnetic Reconnection Inside Flux
  Transfer Event-Like Structures at the Earth's Magnetopause}} {{On the
  Ubiquity of Magnetic Reconnection Inside Flux Transfer Event-Like Structures
  at the Earth's Magnetopause}}.{\BBCQ}
\newblock
\APACjournalVolNumPages{Geophysical Research Letters}{47}{6}{e86726}.
\newblock
\begin{APACrefDOI} \doi{10.1029/2019GL086726} \end{APACrefDOI}
\PrintBackRefs{\CurrentBib}

\bibitem [\protect \citeauthoryear {%
{Fear}%
, {Trenchi}%
, {Coxon}%
\BCBL {}\ \BBA {} {Milan}%
}{%
{Fear}%
\ \protect \BOthers {.}}{%
{\protect \APACyear {2017}}%
}]{%
2017JGRA..12212310F}
\APACinsertmetastar {%
2017JGRA..12212310F}%
\begin{APACrefauthors}%
{Fear}, R\BPBI C.%
, {Trenchi}, L.%
, {Coxon}, J\BPBI C.%
\BCBL {}\ \BBA {} {Milan}, S\BPBI E.%
\end{APACrefauthors}%
\unskip\
\newblock
\APACrefYearMonthDay{2017}{{\APACmonth{12}}}{}.
\newblock
{\BBOQ}\APACrefatitle {{How Much Flux Does a Flux Transfer Event Transfer?}}
  {{How Much Flux Does a Flux Transfer Event Transfer?}}{\BBCQ}
\newblock
\APACjournalVolNumPages{Journal of Geophysical Research (Space
  Physics)}{122}{12}{12,310-12,327}.
\newblock
\begin{APACrefDOI} \doi{10.1002/2017JA024730} \end{APACrefDOI}
\PrintBackRefs{\CurrentBib}

\bibitem [\protect \citeauthoryear {%
{Forbes}%
\ \BBA {} {Lin}%
}{%
{Forbes}%
\ \BBA {} {Lin}%
}{%
{\protect \APACyear {2000}}%
}]{%
Forbes2000}
\APACinsertmetastar {%
Forbes2000}%
\begin{APACrefauthors}%
{Forbes}, T\BPBI G.%
\BCBT {}\ \BBA {} {Lin}, J.%
\end{APACrefauthors}%
\unskip\
\newblock
\APACrefYearMonthDay{2000}{{\APACmonth{11}}}{}.
\newblock
{\BBOQ}\APACrefatitle {{What can we learn about reconnection from coronal mass
  ejections?}} {{What can we learn about reconnection from coronal mass
  ejections?}}{\BBCQ}
\newblock
\APACjournalVolNumPages{Journal of Atmospheric and Solar-Terrestrial
  Physics}{62}{}{1499-1507}.
\newblock
\begin{APACrefDOI} \doi{10.1016/S1364-6826(00)00083-3} \end{APACrefDOI}
\PrintBackRefs{\CurrentBib}

\bibitem [\protect \citeauthoryear {%
{Forbes}%
\ \protect \BOthers {.}}{%
{Forbes}%
\ \protect \BOthers {.}}{%
{\protect \APACyear {2006}}%
}]{%
2006SSRv..123..251F}
\APACinsertmetastar {%
2006SSRv..123..251F}%
\begin{APACrefauthors}%
{Forbes}, T\BPBI G.%
, {Linker}, J\BPBI A.%
, {Chen}, J.%
, {Cid}, C.%
, {K{\'o}ta}, J.%
, {Lee}, M\BPBI A.%
\BDBL {}{Riley}, P.%
\end{APACrefauthors}%
\unskip\
\newblock
\APACrefYearMonthDay{2006}{{\APACmonth{03}}}{}.
\newblock
{\BBOQ}\APACrefatitle {{CME Theory and Models}} {{CME Theory and
  Models}}.{\BBCQ}
\newblock
\APACjournalVolNumPages{\ssr}{123}{1-3}{251-302}.
\newblock
\begin{APACrefDOI} \doi{10.1007/s11214-006-9019-8} \end{APACrefDOI}
\PrintBackRefs{\CurrentBib}

\bibitem [\protect \citeauthoryear {%
{Fu}%
, {Lee}%
\BCBL {}\ \BBA {} {Shi}%
}{%
{Fu}%
\ \protect \BOthers {.}}{%
{\protect \APACyear {1990}}%
}]{%
1990GMS....58..515F}
\APACinsertmetastar {%
1990GMS....58..515F}%
\begin{APACrefauthors}%
{Fu}, Z\BPBI F.%
, {Lee}, L\BPBI C.%
\BCBL {}\ \BBA {} {Shi}, Y.%
\end{APACrefauthors}%
\unskip\
\newblock
\APACrefYearMonthDay{1990}{{\APACmonth{01}}}{}.
\newblock
{\BBOQ}\APACrefatitle {{A three-dimensional MHD simulation of the multiple X
  line reconnection process}} {{A three-dimensional MHD simulation of the
  multiple X line reconnection process}}.{\BBCQ}
\newblock
\APACjournalVolNumPages{Geophysical Monograph Series}{58}{}{515-519}.
\newblock
\begin{APACrefDOI} \doi{10.1029/GM058p0515} \end{APACrefDOI}
\PrintBackRefs{\CurrentBib}

\bibitem [\protect \citeauthoryear {%
{Gosling}%
, {Birn}%
\BCBL {}\ \BBA {} {Hesse}%
}{%
{Gosling}%
\ \protect \BOthers {.}}{%
{\protect \APACyear {1995}}%
}]{%
1995GeoRL..22..869G}
\APACinsertmetastar {%
1995GeoRL..22..869G}%
\begin{APACrefauthors}%
{Gosling}, J\BPBI T.%
, {Birn}, J.%
\BCBL {}\ \BBA {} {Hesse}, M.%
\end{APACrefauthors}%
\unskip\
\newblock
\APACrefYearMonthDay{1995}{{\APACmonth{04}}}{}.
\newblock
{\BBOQ}\APACrefatitle {{Three-dimensional magnetic reconnection and the
  magnetic topology of coronal mass ejection events}} {{Three-dimensional
  magnetic reconnection and the magnetic topology of coronal mass ejection
  events}}.{\BBCQ}
\newblock
\APACjournalVolNumPages{Geophysical Research Letters}{22}{8}{869-872}.
\newblock
\begin{APACrefDOI} \doi{10.1029/95GL00270} \end{APACrefDOI}
\PrintBackRefs{\CurrentBib}

\bibitem [\protect \citeauthoryear {%
{Greenwald}%
\ \protect \BOthers {.}}{%
{Greenwald}%
\ \protect \BOthers {.}}{%
{\protect \APACyear {1995}}%
}]{%
1995SSRv...71..761G}
\APACinsertmetastar {%
1995SSRv...71..761G}%
\begin{APACrefauthors}%
{Greenwald}, R\BPBI A.%
, {Baker}, K\BPBI B.%
, {Dudeney}, J\BPBI R.%
, {Pinnock}, M.%
, {Jones}, T\BPBI B.%
, {Thomas}, E\BPBI C.%
\BDBL {}{Yamagishi}, H.%
\end{APACrefauthors}%
\unskip\
\newblock
\APACrefYearMonthDay{1995}{{\APACmonth{02}}}{}.
\newblock
{\BBOQ}\APACrefatitle {{Darn/Superdarn: A Global View of the Dynamics of
  High-Lattitude Convection}} {{Darn/Superdarn: A Global View of the Dynamics
  of High-Lattitude Convection}}.{\BBCQ}
\newblock
\APACjournalVolNumPages{\ssr}{71}{1-4}{761-796}.
\newblock
\begin{APACrefDOI} \doi{10.1007/BF00751350} \end{APACrefDOI}
\PrintBackRefs{\CurrentBib}

\bibitem [\protect \citeauthoryear {%
{Guo}%
\ \protect \BOthers {.}}{%
{Guo}%
\ \protect \BOthers {.}}{%
{\protect \APACyear {2021}}%
}]{%
2021JGRA..12629388G}
\APACinsertmetastar {%
2021JGRA..12629388G}%
\begin{APACrefauthors}%
{Guo}, J.%
, {Lu}, S.%
, {Lu}, Q.%
, {Lin}, Y.%
, {Wang}, X.%
, {Huang}, K.%
\BDBL {}{Wang}, S.%
\end{APACrefauthors}%
\unskip\
\newblock
\APACrefYearMonthDay{2021}{{\APACmonth{06}}}{}.
\newblock
{\BBOQ}\APACrefatitle {{Re-Reconnection Processes of Magnetopause Flux Ropes:
  Three-Dimensional Global Hybrid Simulations}} {{Re-Reconnection Processes of
  Magnetopause Flux Ropes: Three-Dimensional Global Hybrid
  Simulations}}.{\BBCQ}
\newblock
\APACjournalVolNumPages{Journal of Geophysical Research (Space
  Physics)}{126}{6}{e29388}.
\newblock
\begin{APACrefDOI} \doi{10.1029/2021JA029388} \end{APACrefDOI}
\PrintBackRefs{\CurrentBib}

\bibitem [\protect \citeauthoryear {%
{Hasegawa}%
}{%
{Hasegawa}%
}{%
{\protect \APACyear {2012}}%
}]{%
2012MEEP....1...71H}
\APACinsertmetastar {%
2012MEEP....1...71H}%
\begin{APACrefauthors}%
{Hasegawa}, H.%
\end{APACrefauthors}%
\unskip\
\newblock
\APACrefYearMonthDay{2012}{{\APACmonth{08}}}{}.
\newblock
{\BBOQ}\APACrefatitle {{Structure and Dynamics of the Magnetopause and Its
  Boundary Layers}} {{Structure and Dynamics of the Magnetopause and Its
  Boundary Layers}}.{\BBCQ}
\newblock
\APACjournalVolNumPages{Monographs on Environment, Earth and
  Planets}{1}{2}{71-119}.
\newblock
\begin{APACrefDOI} \doi{10.5047/meep.2012.00102.0071} \end{APACrefDOI}
\PrintBackRefs{\CurrentBib}

\bibitem [\protect \citeauthoryear {%
Hasegawa%
\ \protect \BOthers {.}}{%
Hasegawa%
\ \protect \BOthers {.}}{%
{\protect \APACyear {2007}}%
}]{%
2007JA012492HH}
\APACinsertmetastar {%
2007JA012492HH}%
\begin{APACrefauthors}%
Hasegawa, H.%
, Nakamura, R.%
, Fujimoto, M.%
, Sergeev, V\BPBI A.%
, Lucek, E\BPBI A.%
, Rème, H.%
\BCBL {}\ \BBA {} Khotyaintsev, Y.%
\end{APACrefauthors}%
\unskip\
\newblock
\APACrefYearMonthDay{2007}{}{}.
\newblock
{\BBOQ}\APACrefatitle {Reconstruction of a bipolar magnetic signature in an
  earthward jet in the tail: Flux rope or 3D guide-field reconnection?}
  {Reconstruction of a bipolar magnetic signature in an earthward jet in the
  tail: Flux rope or 3d guide-field reconnection?}{\BBCQ}
\newblock
\APACjournalVolNumPages{Journal of Geophysical Research: Space
  Physics}{112}{A11}{}.
\newblock
\begin{APACrefURL}
  \url{https://agupubs.onlinelibrary.wiley.com/doi/abs/10.1029/2007JA012492}
  \end{APACrefURL}
\newblock
\begin{APACrefDOI} \doi{https://doi.org/10.1029/2007JA012492} \end{APACrefDOI}
\PrintBackRefs{\CurrentBib}

\bibitem [\protect \citeauthoryear {%
{Hasegawa}%
\ \protect \BOthers {.}}{%
{Hasegawa}%
\ \protect \BOthers {.}}{%
{\protect \APACyear {2004}}%
}]{%
Hasegawa2004}
\APACinsertmetastar {%
Hasegawa2004}%
\begin{APACrefauthors}%
{Hasegawa}, H.%
, {Sonnerup}, B.%
, {Dunlop}, M.%
, {Balogh}, A.%
, {Haaland}, S.%
, {Klecker}, B.%
\BDBL {}{R{\`e}me}, H.%
\end{APACrefauthors}%
\unskip\
\newblock
\APACrefYearMonthDay{2004}{{\APACmonth{04}}}{}.
\newblock
{\BBOQ}\APACrefatitle {{Reconstruction of two-dimensional magnetopause
  structures from Cluster observations: verification of method}}
  {{Reconstruction of two-dimensional magnetopause structures from Cluster
  observations: verification of method}}.{\BBCQ}
\newblock
\APACjournalVolNumPages{\anng}{22}{}{1251-1266}.
\newblock
\begin{APACrefDOI} \doi{10.5194/angeo-22-1251-2004} \end{APACrefDOI}
\PrintBackRefs{\CurrentBib}

\bibitem [\protect \citeauthoryear {%
Hasegawa%
\ \protect \BOthers {.}}{%
Hasegawa%
\ \protect \BOthers {.}}{%
{\protect \APACyear {2006}}%
}]{%
angeo-24-603-2006}
\APACinsertmetastar {%
angeo-24-603-2006}%
\begin{APACrefauthors}%
Hasegawa, H.%
, Sonnerup, B\BPBI U\BPBI O.%
, Owen, C\BPBI J.%
, Klecker, B.%
, Paschmann, G.%
, Balogh, A.%
\BCBL {}\ \BBA {} R\`eme, H.%
\end{APACrefauthors}%
\unskip\
\newblock
\APACrefYearMonthDay{2006}{}{}.
\newblock
{\BBOQ}\APACrefatitle {The structure of flux transfer events recovered from
  Cluster data} {The structure of flux transfer events recovered from cluster
  data}.{\BBCQ}
\newblock
\APACjournalVolNumPages{Annales Geophysicae}{24}{2}{603--618}.
\newblock
\begin{APACrefURL} \url{https://angeo.copernicus.org/articles/24/603/2006/}
  \end{APACrefURL}
\newblock
\begin{APACrefDOI} \doi{10.5194/angeo-24-603-2006} \end{APACrefDOI}
\PrintBackRefs{\CurrentBib}

\bibitem [\protect \citeauthoryear {%
{Hasegawa}%
\ \protect \BOthers {.}}{%
{Hasegawa}%
\ \protect \BOthers {.}}{%
{\protect \APACyear {2006}}%
}]{%
2006AnGeo..24..603H}
\APACinsertmetastar {%
2006AnGeo..24..603H}%
\begin{APACrefauthors}%
{Hasegawa}, H.%
, {Sonnerup}, B\BPBI U\BPBI {\"O}.%
, {Owen}, C\BPBI J.%
, {Klecker}, B.%
, {Paschmann}, G.%
, {Balogh}, A.%
\BCBL {}\ \BBA {} {R{\`e}me}, H.%
\end{APACrefauthors}%
\unskip\
\newblock
\APACrefYearMonthDay{2006}{{\APACmonth{03}}}{}.
\newblock
{\BBOQ}\APACrefatitle {{The structure of flux transfer events recovered from
  Cluster data}} {{The structure of flux transfer events recovered from Cluster
  data}}.{\BBCQ}
\newblock
\APACjournalVolNumPages{Annales Geophysicae}{24}{2}{603-618}.
\newblock
\begin{APACrefDOI} \doi{10.5194/angeo-24-603-2006} \end{APACrefDOI}
\PrintBackRefs{\CurrentBib}

\bibitem [\protect \citeauthoryear {%
Hasegawa%
\ \protect \BOthers {.}}{%
Hasegawa%
\ \protect \BOthers {.}}{%
{\protect \APACyear {2010}}%
}]{%
2010GL044219HH}
\APACinsertmetastar {%
2010GL044219HH}%
\begin{APACrefauthors}%
Hasegawa, H.%
, Wang, J.%
, Dunlop, M\BPBI W.%
, Pu, Z\BPBI Y.%
, Zhang, Q\BHBI H.%
, Lavraud, B.%
\BDBL {}Bogdanova, Y\BPBI V.%
\end{APACrefauthors}%
\unskip\
\newblock
\APACrefYearMonthDay{2010}{}{}.
\newblock
{\BBOQ}\APACrefatitle {Evidence for a flux transfer event generated by multiple
  X-line reconnection at the magnetopause} {Evidence for a flux transfer event
  generated by multiple x-line reconnection at the magnetopause}.{\BBCQ}
\newblock
\APACjournalVolNumPages{Geophysical Research Letters}{37}{16}{}.
\newblock
\begin{APACrefURL}
  \url{https://agupubs.onlinelibrary.wiley.com/doi/abs/10.1029/2010GL044219}
  \end{APACrefURL}
\newblock
\begin{APACrefDOI} \doi{https://doi.org/10.1029/2010GL044219} \end{APACrefDOI}
\PrintBackRefs{\CurrentBib}

\bibitem [\protect \citeauthoryear {%
{He}%
, {Hu}%
, {Jiang}%
, {Qiu}%
\BCBL {}\ \BBA {} {Prasad}%
}{%
{He}%
\ \protect \BOthers {.}}{%
{\protect \APACyear {2022}}%
}]{%
he2022quantitative}
\APACinsertmetastar {%
he2022quantitative}%
\begin{APACrefauthors}%
{He}, W.%
, {Hu}, Q.%
, {Jiang}, C.%
, {Qiu}, J.%
\BCBL {}\ \BBA {} {Prasad}, A.%
\end{APACrefauthors}%
\unskip\
\newblock
\APACrefYearMonthDay{2022}{{\APACmonth{08}}}{}.
\newblock
{\BBOQ}\APACrefatitle {{Quantitative Characterization of Magnetic Flux Rope
  Properties for Two Solar Eruption Events}} {{Quantitative Characterization of
  Magnetic Flux Rope Properties for Two Solar Eruption Events}}.{\BBCQ}
\newblock
\APACjournalVolNumPages{\apj}{934}{2}{103}.
\newblock
\begin{APACrefDOI} \doi{10.3847/1538-4357/ac78df} \end{APACrefDOI}
\PrintBackRefs{\CurrentBib}

\bibitem [\protect \citeauthoryear {%
{Hu}%
}{%
{Hu}%
}{%
{\protect \APACyear {2017}}%
}]{%
Hu2017GSreview}
\APACinsertmetastar {%
Hu2017GSreview}%
\begin{APACrefauthors}%
{Hu}, Q.%
\end{APACrefauthors}%
\unskip\
\newblock
\APACrefYearMonthDay{2017}{June}{}.
\newblock
{\BBOQ}\APACrefatitle {{The Grad-Shafranov Reconstruction in Twenty Years: 1996
  - 2016}} {{The Grad-Shafranov Reconstruction in Twenty Years: 1996 -
  2016}}.{\BBCQ}
\newblock
\APACjournalVolNumPages{Sci.~China Earth Sciences}{60}{}{1466-1494}.
\newblock
\begin{APACrefDOI} \doi{doi: 10.1007/s11430-017-9067-2} \end{APACrefDOI}
\PrintBackRefs{\CurrentBib}

\bibitem [\protect \citeauthoryear {%
{Hu}%
, {Qiu}%
, {Dasgupta}%
, {Khare}%
\BCBL {}\ \BBA {} {Webb}%
}{%
{Hu}%
\ \protect \BOthers {.}}{%
{\protect \APACyear {2014}}%
}]{%
2014ApJH}
\APACinsertmetastar {%
2014ApJH}%
\begin{APACrefauthors}%
{Hu}, Q.%
, {Qiu}, J.%
, {Dasgupta}, B.%
, {Khare}, A.%
\BCBL {}\ \BBA {} {Webb}, G\BPBI M.%
\end{APACrefauthors}%
\unskip\
\newblock
\APACrefYearMonthDay{2014}{{\APACmonth{09}}}{}.
\newblock
{\BBOQ}\APACrefatitle {{Structures of Interplanetary Magnetic Flux Ropes and
  Comparison with Their Solar Sources}} {{Structures of Interplanetary Magnetic
  Flux Ropes and Comparison with Their Solar Sources}}.{\BBCQ}
\newblock
\APACjournalVolNumPages{\apj}{793}{}{53}.
\newblock
\begin{APACrefDOI} \doi{10.1088/0004-637X/793/1/53} \end{APACrefDOI}
\PrintBackRefs{\CurrentBib}

\bibitem [\protect \citeauthoryear {%
{Hu}%
, {Qiu}%
\BCBL {}\ \BBA {} {Krucker}%
}{%
{Hu}%
\ \protect \BOthers {.}}{%
{\protect \APACyear {2015}}%
}]{%
2015JGRAH}
\APACinsertmetastar {%
2015JGRAH}%
\begin{APACrefauthors}%
{Hu}, Q.%
, {Qiu}, J.%
\BCBL {}\ \BBA {} {Krucker}, S.%
\end{APACrefauthors}%
\unskip\
\newblock
\APACrefYearMonthDay{2015}{{\APACmonth{06}}}{}.
\newblock
{\BBOQ}\APACrefatitle {{Magnetic field-line lengths inside interplanetary
  magnetic flux ropes}} {{Magnetic field-line lengths inside interplanetary
  magnetic flux ropes}}.{\BBCQ}
\newblock
\APACjournalVolNumPages{\jgr}{120}{}{1}.
\newblock
\begin{APACrefDOI} \doi{10.1002/2015JA021133} \end{APACrefDOI}
\PrintBackRefs{\CurrentBib}

\bibitem [\protect \citeauthoryear {%
{Hu}%
\ \BBA {} {Sonnerup}%
}{%
{Hu}%
\ \BBA {} {Sonnerup}%
}{%
{\protect \APACyear {2002}}%
}]{%
2002JGRAHu}
\APACinsertmetastar {%
2002JGRAHu}%
\begin{APACrefauthors}%
{Hu}, Q.%
\BCBT {}\ \BBA {} {Sonnerup}, B\BPBI U\BPBI {\"O}.%
\end{APACrefauthors}%
\unskip\
\newblock
\APACrefYearMonthDay{2002}{{\APACmonth{07}}}{}.
\newblock
{\BBOQ}\APACrefatitle {{Reconstruction of magnetic clouds in the solar wind:
  Orientations and configurations}} {{Reconstruction of magnetic clouds in the
  solar wind: Orientations and configurations}}.{\BBCQ}
\newblock
\APACjournalVolNumPages{\jgr}{107}{}{1142}.
\newblock
\begin{APACrefDOI} \doi{10.1029/2001JA000293} \end{APACrefDOI}
\PrintBackRefs{\CurrentBib}

\bibitem [\protect \citeauthoryear {%
{Hu}%
\ \protect \BOthers {.}}{%
{Hu}%
\ \protect \BOthers {.}}{%
{\protect \APACyear {2022}}%
}]{%
hu2022validation}
\APACinsertmetastar {%
hu2022validation}%
\begin{APACrefauthors}%
{Hu}, Q.%
, {Zhu}, C.%
, {He}, W.%
, {Qiu}, J.%
, {Jian}, L\BPBI K.%
\BCBL {}\ \BBA {} {Prasad}, A.%
\end{APACrefauthors}%
\unskip\
\newblock
\APACrefYearMonthDay{2022}{{\APACmonth{07}}}{}.
\newblock
{\BBOQ}\APACrefatitle {{Validation and Interpretation of a Three-dimensional
  Configuration of a Magnetic Cloud Flux Rope}} {{Validation and Interpretation
  of a Three-dimensional Configuration of a Magnetic Cloud Flux Rope}}.{\BBCQ}
\newblock
\APACjournalVolNumPages{\apj}{934}{1}{50}.
\newblock
\begin{APACrefDOI} \doi{10.3847/1538-4357/ac7803} \end{APACrefDOI}
\PrintBackRefs{\CurrentBib}

\bibitem [\protect \citeauthoryear {%
Hwang%
\ \protect \BOthers {.}}{%
Hwang%
\ \protect \BOthers {.}}{%
{\protect \APACyear {2020}}%
}]{%
2019JA027674}
\APACinsertmetastar {%
2019JA027674}%
\begin{APACrefauthors}%
Hwang, K\BHBI J.%
, Nishimura, Y.%
, Coster, A\BPBI J.%
, Gillies, R\BPBI G.%
, Fear, R\BPBI C.%
, Fuselier, S\BPBI A.%
\BDBL {}Clausen, L\BPBI B.%
\end{APACrefauthors}%
\unskip\
\newblock
\APACrefYearMonthDay{2020}{}{}.
\newblock
{\BBOQ}\APACrefatitle {Sequential Observations of Flux Transfer Events,
  Poleward-Moving Auroral Forms, and Polar Cap Patches} {Sequential
  observations of flux transfer events, poleward-moving auroral forms, and
  polar cap patches}.{\BBCQ}
\newblock
\APACjournalVolNumPages{Journal of Geophysical Research: Space
  Physics}{125}{6}{e2019JA027674}.
\newblock
\begin{APACrefURL}
  \url{https://agupubs.onlinelibrary.wiley.com/doi/abs/10.1029/2019JA027674}
  \end{APACrefURL}
\newblock
\APACrefnote{e2019JA027674 2019JA027674}
\newblock
\begin{APACrefDOI} \doi{https://doi.org/10.1029/2019JA027674} \end{APACrefDOI}
\PrintBackRefs{\CurrentBib}

\bibitem [\protect \citeauthoryear {%
{Khrabrov}%
\ \BBA {} {Sonnerup}%
}{%
{Khrabrov}%
\ \BBA {} {Sonnerup}%
}{%
{\protect \APACyear {1998}}%
}]{%
1998ISSIRK}
\APACinsertmetastar {%
1998ISSIRK}%
\begin{APACrefauthors}%
{Khrabrov}, A\BPBI V.%
\BCBT {}\ \BBA {} {Sonnerup}, B\BPBI U\BPBI {\"O}.%
\end{APACrefauthors}%
\unskip\
\newblock
\APACrefYearMonthDay{1998}{}{}.
\newblock
{\BBOQ}\APACrefatitle {{DeHoffmann-Teller Analysis}} {{DeHoffmann-Teller
  Analysis}}.{\BBCQ}
\newblock
\APACjournalVolNumPages{ISSI Scientific Reports Series}{1}{}{221-248}.
\PrintBackRefs{\CurrentBib}

\bibitem [\protect \citeauthoryear {%
Kieokaew%
\ \protect \BOthers {.}}{%
Kieokaew%
\ \protect \BOthers {.}}{%
{\protect \APACyear {2021}}%
}]{%
2020GL091257K}
\APACinsertmetastar {%
2020GL091257K}%
\begin{APACrefauthors}%
Kieokaew, R.%
, Lavraud, B.%
, Fargette, N.%
, Marchaudon, A.%
, Génot, V.%
, Jacquey, C.%
\BDBL {}Burch, J.%
\end{APACrefauthors}%
\unskip\
\newblock
\APACrefYearMonthDay{2021}{}{}.
\newblock
{\BBOQ}\APACrefatitle {Statistical Relationship Between Interplanetary Magnetic
  Field Conditions and the Helicity Sign of Flux Transfer Event Flux Ropes}
  {Statistical relationship between interplanetary magnetic field conditions
  and the helicity sign of flux transfer event flux ropes}.{\BBCQ}
\newblock
\APACjournalVolNumPages{Geophysical Research Letters}{48}{6}{e2020GL091257}.
\newblock
\begin{APACrefURL}
  \url{https://agupubs.onlinelibrary.wiley.com/doi/abs/10.1029/2020GL091257}
  \end{APACrefURL}
\newblock
\APACrefnote{e2020GL091257 2020GL091257}
\newblock
\begin{APACrefDOI} \doi{https://doi.org/10.1029/2020GL091257} \end{APACrefDOI}
\PrintBackRefs{\CurrentBib}

\bibitem [\protect \citeauthoryear {%
{Lee}%
\ \BBA {} {Fu}%
}{%
{Lee}%
\ \BBA {} {Fu}%
}{%
{\protect \APACyear {1985}}%
}]{%
1985GeoRL..12..105L}
\APACinsertmetastar {%
1985GeoRL..12..105L}%
\begin{APACrefauthors}%
{Lee}, L\BPBI C.%
\BCBT {}\ \BBA {} {Fu}, Z\BPBI F.%
\end{APACrefauthors}%
\unskip\
\newblock
\APACrefYearMonthDay{1985}{{\APACmonth{02}}}{}.
\newblock
{\BBOQ}\APACrefatitle {{A theory of magnetic flux transfer at the Earth's
  magnetopause}} {{A theory of magnetic flux transfer at the Earth's
  magnetopause}}.{\BBCQ}
\newblock
\APACjournalVolNumPages{Geophysical Research Letters}{12}{2}{105-108}.
\newblock
\begin{APACrefDOI} \doi{10.1029/GL012i002p00105} \end{APACrefDOI}
\PrintBackRefs{\CurrentBib}

\bibitem [\protect \citeauthoryear {%
{Lockwood}%
, {Cowley}%
, {Sandholt}%
\BCBL {}\ \BBA {} {Lepping}%
}{%
{Lockwood}%
\ \protect \BOthers {.}}{%
{\protect \APACyear {1990}}%
}]{%
1990JGR....9517113L}
\APACinsertmetastar {%
1990JGR....9517113L}%
\begin{APACrefauthors}%
{Lockwood}, M.%
, {Cowley}, S\BPBI W\BPBI H.%
, {Sandholt}, P\BPBI E.%
\BCBL {}\ \BBA {} {Lepping}, R\BPBI P.%
\end{APACrefauthors}%
\unskip\
\newblock
\APACrefYearMonthDay{1990}{{\APACmonth{10}}}{}.
\newblock
{\BBOQ}\APACrefatitle {{The ionospheric signatures of flux transfer events and
  solar wind dynamic pressure changes}} {{The ionospheric signatures of flux
  transfer events and solar wind dynamic pressure changes}}.{\BBCQ}
\newblock
\APACjournalVolNumPages{\jgr}{95}{A10}{17113-17135}.
\newblock
\begin{APACrefDOI} \doi{10.1029/JA095iA10p17113} \end{APACrefDOI}
\PrintBackRefs{\CurrentBib}

\bibitem [\protect \citeauthoryear {%
{Longcope}%
\ \protect \BOthers {.}}{%
{Longcope}%
\ \protect \BOthers {.}}{%
{\protect \APACyear {2007}}%
}]{%
Longcope2007b}
\APACinsertmetastar {%
Longcope2007b}%
\begin{APACrefauthors}%
{Longcope}, D.%
, {Beveridge}, C.%
, {Qiu}, J.%
, {Ravindra}, B.%
, {Barnes}, G.%
\BCBL {}\ \BBA {} {Dasso}, S.%
\end{APACrefauthors}%
\unskip\
\newblock
\APACrefYearMonthDay{2007}{{\APACmonth{08}}}{}.
\newblock
{\BBOQ}\APACrefatitle {{Modeling and Measuring the Flux Reconnected and Ejected
  by the Two-Ribbon Flare/CME Event on 7 November 2004}} {{Modeling and
  Measuring the Flux Reconnected and Ejected by the Two-Ribbon Flare/CME Event
  on 7 November 2004}}.{\BBCQ}
\newblock
\APACjournalVolNumPages{\solphys}{244}{}{45-73}.
\newblock
\begin{APACrefDOI} \doi{10.1007/s11207-007-0330-7} \end{APACrefDOI}
\PrintBackRefs{\CurrentBib}

\bibitem [\protect \citeauthoryear {%
{Marchaudon}%
, {Cerisier}%
, {Bosqued}%
\BCBL {}\ \protect \BOthers {.}}{%
{Marchaudon}%
, {Cerisier}%
, {Bosqued}%
\BCBL {}\ \protect \BOthers {.}}{%
{\protect \APACyear {2004}}%
}]{%
2004AnGeo..22..141M}
\APACinsertmetastar {%
2004AnGeo..22..141M}%
\begin{APACrefauthors}%
{Marchaudon}, A.%
, {Cerisier}, J.%
, {Bosqued}, J.%
, {Dunlop}, M.%
, {Wild}, J.%
, {D{\'e}cr{\'e}au}, P.%
\BDBL {}{Laakso}, H.%
\end{APACrefauthors}%
\unskip\
\newblock
\APACrefYearMonthDay{2004}{{\APACmonth{01}}}{}.
\newblock
{\BBOQ}\APACrefatitle {{Transient plasma injections in the dayside
  magnetosphere: one-to-one correlated observations by Cluster and SuperDARN}}
  {{Transient plasma injections in the dayside magnetosphere: one-to-one
  correlated observations by Cluster and SuperDARN}}.{\BBCQ}
\newblock
\APACjournalVolNumPages{Annales Geophysicae}{22}{1}{141-158}.
\newblock
\begin{APACrefDOI} \doi{10.5194/angeo-22-141-2004} \end{APACrefDOI}
\PrintBackRefs{\CurrentBib}

\bibitem [\protect \citeauthoryear {%
{Marchaudon}%
, {Cerisier}%
, {Greenwald}%
\BCBL {}\ \BBA {} {Sofko}%
}{%
{Marchaudon}%
, {Cerisier}%
, {Greenwald}%
\BCBL {}\ \BBA {} {Sofko}%
}{%
{\protect \APACyear {2004}}%
}]{%
2004GeoRL..31.9809M}
\APACinsertmetastar {%
2004GeoRL..31.9809M}%
\begin{APACrefauthors}%
{Marchaudon}, A.%
, {Cerisier}, J\BPBI C.%
, {Greenwald}, R\BPBI A.%
\BCBL {}\ \BBA {} {Sofko}, G\BPBI J.%
\end{APACrefauthors}%
\unskip\
\newblock
\APACrefYearMonthDay{2004}{{\APACmonth{05}}}{}.
\newblock
{\BBOQ}\APACrefatitle {{Electrodynamics of a flux transfer event: Experimental
  test of the Southwood model}} {{Electrodynamics of a flux transfer event:
  Experimental test of the Southwood model}}.{\BBCQ}
\newblock
\APACjournalVolNumPages{Geophysical Research Letters}{31}{9}{L09809}.
\newblock
\begin{APACrefDOI} \doi{10.1029/2004GL019922} \end{APACrefDOI}
\PrintBackRefs{\CurrentBib}

\bibitem [\protect \citeauthoryear {%
Martin%
\ \protect \BOthers {.}}{%
Martin%
\ \protect \BOthers {.}}{%
{\protect \APACyear {2023}}%
}]{%
superdarn_data_visualization_working_gro_2023_7767590}
\APACinsertmetastar {%
superdarn_data_visualization_working_gro_2023_7767590}%
\begin{APACrefauthors}%
Martin, C.%
, Shi, X.%
, Schmidt, M.%
, Day, E\BPBI K.%
, Bland, E.%
, Khanal, K.%
\BDBL {}Krieger, K.%
\end{APACrefauthors}%
\unskip\
\newblock
\APACrefYearMonthDay{2023}{{\APACmonth{03}}}{}.
\newblock
\APACrefbtitle {SuperDARN/pydarn: pyDARN v3.1.1.} {Superdarn/pydarn: pydarn
  v3.1.1.}
\newblock
\APACaddressPublisher{}{Zenodo}.
\newblock
\begin{APACrefURL} \url{https://doi.org/10.5281/zenodo.7767590}
  \end{APACrefURL}
\newblock
\begin{APACrefDOI} \doi{10.5281/zenodo.7767590} \end{APACrefDOI}
\PrintBackRefs{\CurrentBib}

\bibitem [\protect \citeauthoryear {%
{Milan}%
, {Lester}%
, {Cowley}%
\BCBL {}\ \BBA {} {Brittnacher}%
}{%
{Milan}%
\ \protect \BOthers {.}}{%
{\protect \APACyear {2000}}%
}]{%
2000JGR...10515741M}
\APACinsertmetastar {%
2000JGR...10515741M}%
\begin{APACrefauthors}%
{Milan}, S\BPBI E.%
, {Lester}, M.%
, {Cowley}, S\BPBI W\BPBI H.%
\BCBL {}\ \BBA {} {Brittnacher}, M.%
\end{APACrefauthors}%
\unskip\
\newblock
\APACrefYearMonthDay{2000}{{\APACmonth{07}}}{}.
\newblock
{\BBOQ}\APACrefatitle {{Convection and auroral response to a southward turning
  of the IMF: Polar UVI, CUTLASS, and IMAGE signatures of transient magnetic
  flux transfer at the magnetopause}} {{Convection and auroral response to a
  southward turning of the IMF: Polar UVI, CUTLASS, and IMAGE signatures of
  transient magnetic flux transfer at the magnetopause}}.{\BBCQ}
\newblock
\APACjournalVolNumPages{\jgr}{105}{A7}{15741-15756}.
\newblock
\begin{APACrefDOI} \doi{10.1029/2000JA900022} \end{APACrefDOI}
\PrintBackRefs{\CurrentBib}

\bibitem [\protect \citeauthoryear {%
{Nishitani}%
\ \protect \BOthers {.}}{%
{Nishitani}%
\ \protect \BOthers {.}}{%
{\protect \APACyear {2019}}%
}]{%
2019PEPS....6...27N}
\APACinsertmetastar {%
2019PEPS....6...27N}%
\begin{APACrefauthors}%
{Nishitani}, N.%
, {Ruohoniemi}, J\BPBI M.%
, {Lester}, M.%
, {Baker}, J\BPBI B\BPBI H.%
, {Koustov}, A\BPBI V.%
, {Shepherd}, S\BPBI G.%
\BDBL {}{Kikuchi}, T.%
\end{APACrefauthors}%
\unskip\
\newblock
\APACrefYearMonthDay{2019}{{\APACmonth{03}}}{}.
\newblock
{\BBOQ}\APACrefatitle {{Review of the accomplishments of mid-latitude Super
  Dual Auroral Radar Network (SuperDARN) HF radars}} {{Review of the
  accomplishments of mid-latitude Super Dual Auroral Radar Network (SuperDARN)
  HF radars}}.{\BBCQ}
\newblock
\APACjournalVolNumPages{Progress in Earth and Planetary Science}{6}{1}{27}.
\newblock
\begin{APACrefDOI} \doi{10.1186/s40645-019-0270-5} \end{APACrefDOI}
\PrintBackRefs{\CurrentBib}

\bibitem [\protect \citeauthoryear {%
{Oksavik}%
\ \protect \BOthers {.}}{%
{Oksavik}%
\ \protect \BOthers {.}}{%
{\protect \APACyear {2005}}%
}]{%
2005AnGeo..23.2657O}
\APACinsertmetastar {%
2005AnGeo..23.2657O}%
\begin{APACrefauthors}%
{Oksavik}, K.%
, {Moen}, J.%
, {Carlson}, H\BPBI C.%
, {Greenwald}, R\BPBI A.%
, {Milan}, S\BPBI E.%
, {Lester}, M.%
\BDBL {}{Barnes}, R\BPBI J.%
\end{APACrefauthors}%
\unskip\
\newblock
\APACrefYearMonthDay{2005}{{\APACmonth{10}}}{}.
\newblock
{\BBOQ}\APACrefatitle {{Multi-instrument mapping of the small-scale flow
  dynamics related to a cusp auroral transient}} {{Multi-instrument mapping of
  the small-scale flow dynamics related to a cusp auroral transient}}.{\BBCQ}
\newblock
\APACjournalVolNumPages{Annales Geophysicae}{23}{7}{2657-2670}.
\newblock
\begin{APACrefDOI} \doi{10.5194/angeo-23-2657-2005} \end{APACrefDOI}
\PrintBackRefs{\CurrentBib}

\bibitem [\protect \citeauthoryear {%
{Paschmann}%
\ \BBA {} {Sonnerup}%
}{%
{Paschmann}%
\ \BBA {} {Sonnerup}%
}{%
{\protect \APACyear {2008}}%
}]{%
2008ISSIR...8...65P}
\APACinsertmetastar {%
2008ISSIR...8...65P}%
\begin{APACrefauthors}%
{Paschmann}, G.%
\BCBT {}\ \BBA {} {Sonnerup}, B\BPBI U\BPBI O.%
\end{APACrefauthors}%
\unskip\
\newblock
\APACrefYearMonthDay{2008}{}{}.
\newblock
{\BBOQ}\APACrefatitle {{Proper Frame Determination and Walen Test}} {{Proper
  Frame Determination and Walen Test}}.{\BBCQ}
\newblock
\APACjournalVolNumPages{ISSI Scientific Reports Series}{8}{}{65-74}.
\PrintBackRefs{\CurrentBib}

\bibitem [\protect \citeauthoryear {%
Phan%
, Paschmann%
, Baumjohann%
, Sckopke%
\BCBL {}\ \BBA {} Lühr%
}{%
Phan%
\ \protect \BOthers {.}}{%
{\protect \APACyear {1994}}%
}]{%
Phan93JA02444}
\APACinsertmetastar {%
Phan93JA02444}%
\begin{APACrefauthors}%
Phan, T\BPBI D.%
, Paschmann, G.%
, Baumjohann, W.%
, Sckopke, N.%
\BCBL {}\ \BBA {} Lühr, H.%
\end{APACrefauthors}%
\unskip\
\newblock
\APACrefYearMonthDay{1994}{}{}.
\newblock
{\BBOQ}\APACrefatitle {The magnetosheath region adjacent to the dayside
  magnetopause: AMPTE/IRM observations} {The magnetosheath region adjacent to
  the dayside magnetopause: Ampte/irm observations}.{\BBCQ}
\newblock
\APACjournalVolNumPages{Journal of Geophysical Research: Space
  Physics}{99}{A1}{121-141}.
\newblock
\begin{APACrefURL}
  \url{https://agupubs.onlinelibrary.wiley.com/doi/abs/10.1029/93JA02444}
  \end{APACrefURL}
\newblock
\begin{APACrefDOI} \doi{https://doi.org/10.1029/93JA02444} \end{APACrefDOI}
\PrintBackRefs{\CurrentBib}

\bibitem [\protect \citeauthoryear {%
{Pollock}%
\ \protect \BOthers {.}}{%
{Pollock}%
\ \protect \BOthers {.}}{%
{\protect \APACyear {2016}}%
}]{%
2016SSRv..199..331P}
\APACinsertmetastar {%
2016SSRv..199..331P}%
\begin{APACrefauthors}%
{Pollock}, C.%
, {Moore}, T.%
, {Jacques}, A.%
, {Burch}, J.%
, {Gliese}, U.%
, {Saito}, Y.%
\BDBL {}{Zeuch}, M.%
\end{APACrefauthors}%
\unskip\
\newblock
\APACrefYearMonthDay{2016}{{\APACmonth{03}}}{}.
\newblock
{\BBOQ}\APACrefatitle {{Fast Plasma Investigation for Magnetospheric
  Multiscale}} {{Fast Plasma Investigation for Magnetospheric
  Multiscale}}.{\BBCQ}
\newblock
\APACjournalVolNumPages{\ssr}{199}{1-4}{331-406}.
\newblock
\begin{APACrefDOI} \doi{10.1007/s11214-016-0245-4} \end{APACrefDOI}
\PrintBackRefs{\CurrentBib}

\bibitem [\protect \citeauthoryear {%
Qi%
, Russell%
, Jia%
\BCBL {}\ \BBA {} Hubbert%
}{%
Qi%
\ \protect \BOthers {.}}{%
{\protect \APACyear {2020}}%
}]{%
2020GL090314entanglement}
\APACinsertmetastar {%
2020GL090314entanglement}%
\begin{APACrefauthors}%
Qi, Y.%
, Russell, C\BPBI T.%
, Jia, Y\BHBI D.%
\BCBL {}\ \BBA {} Hubbert, M.%
\end{APACrefauthors}%
\unskip\
\newblock
\APACrefYearMonthDay{2020}{}{}.
\newblock
{\BBOQ}\APACrefatitle {Temporal Evolution of Flux Tube Entanglement at the
  Magnetopause as Observed by the MMS Satellites} {Temporal evolution of flux
  tube entanglement at the magnetopause as observed by the mms
  satellites}.{\BBCQ}
\newblock
\APACjournalVolNumPages{Geophysical Research Letters}{47}{23}{e2020GL090314}.
\newblock
\begin{APACrefURL}
  \url{https://agupubs.onlinelibrary.wiley.com/doi/abs/10.1029/2020GL090314}
  \end{APACrefURL}
\newblock
\APACrefnote{e2020GL090314 2020GL090314}
\newblock
\begin{APACrefDOI} \doi{https://doi.org/10.1029/2020GL090314} \end{APACrefDOI}
\PrintBackRefs{\CurrentBib}

\bibitem [\protect \citeauthoryear {%
{Qiu}%
, {Hu}%
, {Howard}%
\BCBL {}\ \BBA {} {Yurchyshyn}%
}{%
{Qiu}%
\ \protect \BOthers {.}}{%
{\protect \APACyear {2007}}%
}]{%
Qiu2007}
\APACinsertmetastar {%
Qiu2007}%
\begin{APACrefauthors}%
{Qiu}, J.%
, {Hu}, Q.%
, {Howard}, T\BPBI A.%
\BCBL {}\ \BBA {} {Yurchyshyn}, V\BPBI B.%
\end{APACrefauthors}%
\unskip\
\newblock
\APACrefYearMonthDay{2007}{{\APACmonth{04}}}{}.
\newblock
{\BBOQ}\APACrefatitle {{On the Magnetic Flux Budget in Low-Corona Magnetic
  Reconnection and Interplanetary Coronal Mass Ejections}} {{On the Magnetic
  Flux Budget in Low-Corona Magnetic Reconnection and Interplanetary Coronal
  Mass Ejections}}.{\BBCQ}
\newblock
\APACjournalVolNumPages{\apj}{659}{}{758-772}.
\newblock
\begin{APACrefDOI} \doi{10.1086/512060} \end{APACrefDOI}
\PrintBackRefs{\CurrentBib}

\bibitem [\protect \citeauthoryear {%
{Raeder}%
}{%
{Raeder}%
}{%
{\protect \APACyear {2006}}%
}]{%
2006AnGeo..24..381R}
\APACinsertmetastar {%
2006AnGeo..24..381R}%
\begin{APACrefauthors}%
{Raeder}, J.%
\end{APACrefauthors}%
\unskip\
\newblock
\APACrefYearMonthDay{2006}{{\APACmonth{03}}}{}.
\newblock
{\BBOQ}\APACrefatitle {{Flux Transfer Events: 1. generation mechanism for
  strong southward IMF}} {{Flux Transfer Events: 1. generation mechanism for
  strong southward IMF}}.{\BBCQ}
\newblock
\APACjournalVolNumPages{Annales Geophysicae}{24}{1}{381-392}.
\newblock
\begin{APACrefDOI} \doi{10.5194/angeo-24-381-2006} \end{APACrefDOI}
\PrintBackRefs{\CurrentBib}

\bibitem [\protect \citeauthoryear {%
{Russell}%
\ \protect \BOthers {.}}{%
{Russell}%
\ \protect \BOthers {.}}{%
{\protect \APACyear {2016}}%
}]{%
2016SSRv..199..189R}
\APACinsertmetastar {%
2016SSRv..199..189R}%
\begin{APACrefauthors}%
{Russell}, C\BPBI T.%
, {Anderson}, B\BPBI J.%
, {Baumjohann}, W.%
, {Bromund}, K\BPBI R.%
, {Dearborn}, D.%
, {Fischer}, D.%
\BDBL {}{Richter}, I.%
\end{APACrefauthors}%
\unskip\
\newblock
\APACrefYearMonthDay{2016}{{\APACmonth{03}}}{}.
\newblock
{\BBOQ}\APACrefatitle {{The Magnetospheric Multiscale Magnetometers}} {{The
  Magnetospheric Multiscale Magnetometers}}.{\BBCQ}
\newblock
\APACjournalVolNumPages{\ssr}{199}{1-4}{189-256}.
\newblock
\begin{APACrefDOI} \doi{10.1007/s11214-014-0057-3} \end{APACrefDOI}
\PrintBackRefs{\CurrentBib}

\bibitem [\protect \citeauthoryear {%
{Russell}%
\ \BBA {} {Elphic}%
}{%
{Russell}%
\ \BBA {} {Elphic}%
}{%
{\protect \APACyear {1978}}%
}]{%
1978SSRv...22..681R}
\APACinsertmetastar {%
1978SSRv...22..681R}%
\begin{APACrefauthors}%
{Russell}, C\BPBI T.%
\BCBT {}\ \BBA {} {Elphic}, R\BPBI C.%
\end{APACrefauthors}%
\unskip\
\newblock
\APACrefYearMonthDay{1978}{{\APACmonth{12}}}{}.
\newblock
{\BBOQ}\APACrefatitle {{Initial ISEE Magnetometer Results: Magnetopause
  Observations (Article published in the special issues: Advances in
  Magnetospheric Physics with GEOS- 1 and ISEE - 1 and 2.)}} {{Initial ISEE
  Magnetometer Results: Magnetopause Observations (Article published in the
  special issues: Advances in Magnetospheric Physics with GEOS- 1 and ISEE - 1
  and 2.)}}.{\BBCQ}
\newblock
\APACjournalVolNumPages{\ssr}{22}{6}{681-715}.
\newblock
\begin{APACrefDOI} \doi{10.1007/BF00212619} \end{APACrefDOI}
\PrintBackRefs{\CurrentBib}

\bibitem [\protect \citeauthoryear {%
{Russell}%
, {Priest}%
\BCBL {}\ \BBA {} {Lee}%
}{%
{Russell}%
\ \protect \BOthers {.}}{%
{\protect \APACyear {1990}}%
}]{%
1990GMS....58.....R}
\APACinsertmetastar {%
1990GMS....58.....R}%
\begin{APACrefauthors}%
{Russell}, C\BPBI T.%
, {Priest}, E\BPBI R.%
\BCBL {}\ \BBA {} {Lee}, L\BPBI C.%
\end{APACrefauthors}%
\unskip\
\newblock
\APACrefYearMonthDay{1990}{{\APACmonth{01}}}{}.
\newblock
{\BBOQ}\APACrefatitle {{Physics of magnetic flux ropes}} {{Physics of magnetic
  flux ropes}}.{\BBCQ}
\newblock
\APACjournalVolNumPages{Washington DC American Geophysical Union Geophysical
  Monograph Series}{58}{}{}.
\newblock
\begin{APACrefDOI} \doi{10.1029/GM058} \end{APACrefDOI}
\PrintBackRefs{\CurrentBib}

\bibitem [\protect \citeauthoryear {%
{Sandholt}%
\ \protect \BOthers {.}}{%
{Sandholt}%
\ \protect \BOthers {.}}{%
{\protect \APACyear {1986}}%
}]{%
1986JGR....9110063S}
\APACinsertmetastar {%
1986JGR....9110063S}%
\begin{APACrefauthors}%
{Sandholt}, P\BPBI E.%
, {Deehr}, C\BPBI S.%
, {Egeland}, A.%
, {Lybekk}, B.%
, {Viereck}, R.%
\BCBL {}\ \BBA {} {Romick}, G\BPBI J.%
\end{APACrefauthors}%
\unskip\
\newblock
\APACrefYearMonthDay{1986}{{\APACmonth{09}}}{}.
\newblock
{\BBOQ}\APACrefatitle {{Signatures in the dayside aurora of plasma transfer
  from the magnetosheath}} {{Signatures in the dayside aurora of plasma
  transfer from the magnetosheath}}.{\BBCQ}
\newblock
\APACjournalVolNumPages{\jgr}{91}{A9}{10063-10079}.
\newblock
\begin{APACrefDOI} \doi{10.1029/JA091iA09p10063} \end{APACrefDOI}
\PrintBackRefs{\CurrentBib}

\bibitem [\protect \citeauthoryear {%
Shepherd%
}{%
Shepherd%
}{%
{\protect \APACyear {2014}}%
}]{%
2014JA020264Shepherd}
\APACinsertmetastar {%
2014JA020264Shepherd}%
\begin{APACrefauthors}%
Shepherd, S\BPBI G.%
\end{APACrefauthors}%
\unskip\
\newblock
\APACrefYearMonthDay{2014}{}{}.
\newblock
{\BBOQ}\APACrefatitle {Altitude-adjusted corrected geomagnetic coordinates:
  Definition and functional approximations} {Altitude-adjusted corrected
  geomagnetic coordinates: Definition and functional approximations}.{\BBCQ}
\newblock
\APACjournalVolNumPages{Journal of Geophysical Research: Space
  Physics}{119}{9}{7501-7521}.
\newblock
\begin{APACrefURL}
  \url{https://agupubs.onlinelibrary.wiley.com/doi/abs/10.1002/2014JA020264}
  \end{APACrefURL}
\newblock
\begin{APACrefDOI} \doi{https://doi.org/10.1002/2014JA020264} \end{APACrefDOI}
\PrintBackRefs{\CurrentBib}

\bibitem [\protect \citeauthoryear {%
Shi%
\ \protect \BOthers {.}}{%
Shi%
\ \protect \BOthers {.}}{%
{\protect \APACyear {2022}}%
}]{%
Shifspas.2022.1022690}
\APACinsertmetastar {%
Shifspas.2022.1022690}%
\begin{APACrefauthors}%
Shi, X.%
, Schmidt, M.%
, Martin, C\BPBI J.%
, Billett, D\BPBI D.%
, Bland, E.%
, Tholley, F\BPBI H.%
\BDBL {}McWilliams, K.%
\end{APACrefauthors}%
\unskip\
\newblock
\APACrefYearMonthDay{2022}{}{}.
\newblock
{\BBOQ}\APACrefatitle {pyDARN: A Python software for visualizing SuperDARN
  radar data} {pydarn: A python software for visualizing superdarn radar
  data}.{\BBCQ}
\newblock
\APACjournalVolNumPages{Frontiers in Astronomy and Space Sciences}{9}{}{}.
\newblock
\begin{APACrefURL}
  \url{https://www.frontiersin.org/articles/10.3389/fspas.2022.1022690}
  \end{APACrefURL}
\newblock
\begin{APACrefDOI} \doi{10.3389/fspas.2022.1022690} \end{APACrefDOI}
\PrintBackRefs{\CurrentBib}

\bibitem [\protect \citeauthoryear {%
{Shue}%
\ \protect \BOthers {.}}{%
{Shue}%
\ \protect \BOthers {.}}{%
{\protect \APACyear {1998}}%
}]{%
1998JGR...10317691S}
\APACinsertmetastar {%
1998JGR...10317691S}%
\begin{APACrefauthors}%
{Shue}, J\BPBI H.%
, {Song}, P.%
, {Russell}, C\BPBI T.%
, {Steinberg}, J\BPBI T.%
, {Chao}, J\BPBI K.%
, {Zastenker}, G.%
\BDBL {}{Kawano}, H.%
\end{APACrefauthors}%
\unskip\
\newblock
\APACrefYearMonthDay{1998}{{\APACmonth{08}}}{}.
\newblock
{\BBOQ}\APACrefatitle {{Magnetopause location under extreme solar wind
  conditions}} {{Magnetopause location under extreme solar wind
  conditions}}.{\BBCQ}
\newblock
\APACjournalVolNumPages{\jgr}{103}{A8}{17691-17700}.
\newblock
\begin{APACrefDOI} \doi{10.1029/98JA01103} \end{APACrefDOI}
\PrintBackRefs{\CurrentBib}

\bibitem [\protect \citeauthoryear {%
{Sonnerup}%
, {Hasegawa}%
\BCBL {}\ \BBA {} {Paschmann}%
}{%
{Sonnerup}%
\ \protect \BOthers {.}}{%
{\protect \APACyear {2004}}%
}]{%
Sonnerup2004}
\APACinsertmetastar {%
Sonnerup2004}%
\begin{APACrefauthors}%
{Sonnerup}, B\BPBI U\BPBI {\"O}.%
, {Hasegawa}, H.%
\BCBL {}\ \BBA {} {Paschmann}, G.%
\end{APACrefauthors}%
\unskip\
\newblock
\APACrefYearMonthDay{2004}{{\APACmonth{06}}}{}.
\newblock
{\BBOQ}\APACrefatitle {{Anatomy of a flux transfer event seen by Cluster}}
  {{Anatomy of a flux transfer event seen by Cluster}}.{\BBCQ}
\newblock
\APACjournalVolNumPages{Geophysical Research Letters}{31}{}{11803}.
\newblock
\begin{APACrefDOI} \doi{10.1029/2004GL020134} \end{APACrefDOI}
\PrintBackRefs{\CurrentBib}

\bibitem [\protect \citeauthoryear {%
Thomas%
\ \BBA {} Shepherd%
}{%
Thomas%
\ \BBA {} Shepherd%
}{%
{\protect \APACyear {2018}}%
}]{%
2018JA025280}
\APACinsertmetastar {%
2018JA025280}%
\begin{APACrefauthors}%
Thomas, E\BPBI G.%
\BCBT {}\ \BBA {} Shepherd, S\BPBI G.%
\end{APACrefauthors}%
\unskip\
\newblock
\APACrefYearMonthDay{2018}{}{}.
\newblock
{\BBOQ}\APACrefatitle {Statistical Patterns of Ionospheric Convection Derived
  From Mid-latitude, High-Latitude, and Polar SuperDARN HF Radar Observations}
  {Statistical patterns of ionospheric convection derived from mid-latitude,
  high-latitude, and polar superdarn hf radar observations}.{\BBCQ}
\newblock
\APACjournalVolNumPages{Journal of Geophysical Research: Space
  Physics}{123}{4}{3196-3216}.
\newblock
\begin{APACrefURL}
  \url{https://agupubs.onlinelibrary.wiley.com/doi/abs/10.1002/2018JA025280}
  \end{APACrefURL}
\newblock
\begin{APACrefDOI} \doi{https://doi.org/10.1002/2018JA025280} \end{APACrefDOI}
\PrintBackRefs{\CurrentBib}

\bibitem [\protect \citeauthoryear {%
{Tsyganenko}%
}{%
{Tsyganenko}%
}{%
{\protect \APACyear {1996}}%
}]{%
1996ESASP.389..181T}
\APACinsertmetastar {%
1996ESASP.389..181T}%
\begin{APACrefauthors}%
{Tsyganenko}, N\BPBI A.%
\end{APACrefauthors}%
\unskip\
\newblock
\APACrefYearMonthDay{1996}{{\APACmonth{10}}}{}.
\newblock
{\BBOQ}\APACrefatitle {{Effects of the solar wind conditions in the global
  magnetospheric configurations as deduced from data-based field models
  (Invited)}} {{Effects of the solar wind conditions in the global
  magnetospheric configurations as deduced from data-based field models
  (Invited)}}.{\BBCQ}
\newblock
\BIn{} E\BPBI J.~{Rolfe}\ \BBA {} B.~{Kaldeich}\ (\BEDS), \APACrefbtitle
  {International Conference on Substorms} {International conference on
  substorms}\ (\BVOL~389, \BPG~181).
\PrintBackRefs{\CurrentBib}

\bibitem [\protect \citeauthoryear {%
{Vorobev}%
, {Starkov}%
, {Gustafsson}%
, {Feldshtein}%
\BCBL {}\ \BBA {} {Shevnina}%
}{%
{Vorobev}%
\ \protect \BOthers {.}}{%
{\protect \APACyear {1975}}%
}]{%
1975P&SS...23..269V}
\APACinsertmetastar {%
1975P&SS...23..269V}%
\begin{APACrefauthors}%
{Vorobev}, V\BPBI G.%
, {Starkov}, G\BPBI V.%
, {Gustafsson}, G.%
, {Feldshtein}, I\BPBI I.%
\BCBL {}\ \BBA {} {Shevnina}, N\BPBI F.%
\end{APACrefauthors}%
\unskip\
\newblock
\APACrefYearMonthDay{1975}{{\APACmonth{02}}}{}.
\newblock
{\BBOQ}\APACrefatitle {{Dynamics of day and night aurora during substorms}}
  {{Dynamics of day and night aurora during substorms}}.{\BBCQ}
\newblock
\APACjournalVolNumPages{\planss}{23}{2}{269-278}.
\newblock
\begin{APACrefDOI} \doi{10.1016/0032-0633(75)90132-4} \end{APACrefDOI}
\PrintBackRefs{\CurrentBib}

\bibitem [\protect \citeauthoryear {%
{Wild}%
\ \protect \BOthers {.}}{%
{Wild}%
\ \protect \BOthers {.}}{%
{\protect \APACyear {2003}}%
}]{%
2003AnGeo..21.1807W}
\APACinsertmetastar {%
2003AnGeo..21.1807W}%
\begin{APACrefauthors}%
{Wild}, J\BPBI A.%
, {Milan}, S\BPBI E.%
, {Cowley}, S\BPBI W\BPBI H.%
, {Dunlop}, M\BPBI W.%
, {Owen}, C\BPBI J.%
, {Bosqued}, J\BPBI M.%
\BDBL {}{R{\`e}me}, H.%
\end{APACrefauthors}%
\unskip\
\newblock
\APACrefYearMonthDay{2003}{{\APACmonth{08}}}{}.
\newblock
{\BBOQ}\APACrefatitle {{Coordinated interhemispheric SuperDARN radar
  observations of the ionospheric response to flux transfer events observed by
  the Cluster spacecraft at the high-latitude magnetopause}} {{Coordinated
  interhemispheric SuperDARN radar observations of the ionospheric response to
  flux transfer events observed by the Cluster spacecraft at the high-latitude
  magnetopause}}.{\BBCQ}
\newblock
\APACjournalVolNumPages{Annales Geophysicae}{21}{8}{1807-1826}.
\newblock
\begin{APACrefDOI} \doi{10.5194/angeo-21-1807-2003} \end{APACrefDOI}
\PrintBackRefs{\CurrentBib}

\bibitem [\protect \citeauthoryear {%
{Zhang}%
\ \protect \BOthers {.}}{%
{Zhang}%
\ \protect \BOthers {.}}{%
{\protect \APACyear {2022}}%
}]{%
2022SSRv..218...40Z}
\APACinsertmetastar {%
2022SSRv..218...40Z}%
\begin{APACrefauthors}%
{Zhang}, H.%
, {Zong}, Q.%
, {Connor}, H.%
, {Delamere}, P.%
, {Facsk{\'o}}, G.%
, {Han}, D.%
\BDBL {}{Yao}, S.%
\end{APACrefauthors}%
\unskip\
\newblock
\APACrefYearMonthDay{2022}{{\APACmonth{08}}}{}.
\newblock
{\BBOQ}\APACrefatitle {{Dayside Transient Phenomena and Their Impact on the
  Magnetosphere and Ionosphere}} {{Dayside Transient Phenomena and Their Impact
  on the Magnetosphere and Ionosphere}}.{\BBCQ}
\newblock
\APACjournalVolNumPages{\ssr}{218}{5}{40}.
\newblock
\begin{APACrefDOI} \doi{10.1007/s11214-021-00865-0} \end{APACrefDOI}
\PrintBackRefs{\CurrentBib}

\bibitem [\protect \citeauthoryear {%
Zou%
\ \protect \BOthers {.}}{%
Zou%
\ \protect \BOthers {.}}{%
{\protect \APACyear {2022}}%
}]{%
2021GL096583Zou}
\APACinsertmetastar {%
2021GL096583Zou}%
\begin{APACrefauthors}%
Zou, Y.%
, Walsh, B\BPBI M.%
, Chen, L\BHBI J.%
, Ng, J.%
, Shi, X.%
, Wang, C\BHBI P.%
\BDBL {}Michael~Ruohoniemi, J.%
\end{APACrefauthors}%
\unskip\
\newblock
\APACrefYearMonthDay{2022}{}{}.
\newblock
{\BBOQ}\APACrefatitle {Unsteady Magnetopause Reconnection Under Quasi-Steady
  Solar Wind Driving} {Unsteady magnetopause reconnection under quasi-steady
  solar wind driving}.{\BBCQ}
\newblock
\APACjournalVolNumPages{Geophysical Research Letters}{49}{1}{e2021GL096583}.
\newblock
\begin{APACrefURL}
  \url{https://agupubs.onlinelibrary.wiley.com/doi/abs/10.1029/2021GL096583}
  \end{APACrefURL}
\newblock
\APACrefnote{e2021GL096583 2021GL096583}
\newblock
\begin{APACrefDOI} \doi{https://doi.org/10.1029/2021GL096583} \end{APACrefDOI}
\PrintBackRefs{\CurrentBib}

\bibitem [\protect \citeauthoryear {%
Zou%
\ \protect \BOthers {.}}{%
Zou%
\ \protect \BOthers {.}}{%
{\protect \APACyear {2018}}%
}]{%
2017GL075765Zou}
\APACinsertmetastar {%
2017GL075765Zou}%
\begin{APACrefauthors}%
Zou, Y.%
, Walsh, B\BPBI M.%
, Nishimura, Y.%
, Angelopoulos, V.%
, Ruohoniemi, J\BPBI M.%
, McWilliams, K\BPBI A.%
\BCBL {}\ \BBA {} Nishitani, N.%
\end{APACrefauthors}%
\unskip\
\newblock
\APACrefYearMonthDay{2018}{}{}.
\newblock
{\BBOQ}\APACrefatitle {Spreading Speed of Magnetopause Reconnection X-Lines
  Using Ground-Satellite Coordination} {Spreading speed of magnetopause
  reconnection x-lines using ground-satellite coordination}.{\BBCQ}
\newblock
\APACjournalVolNumPages{Geophysical Research Letters}{45}{1}{80-89}.
\newblock
\begin{APACrefURL}
  \url{https://agupubs.onlinelibrary.wiley.com/doi/abs/10.1002/2017GL075765}
  \end{APACrefURL}
\newblock
\begin{APACrefDOI} \doi{https://doi.org/10.1002/2017GL075765} \end{APACrefDOI}
\PrintBackRefs{\CurrentBib}

\bibitem [\protect \citeauthoryear {%
Zou%
\ \protect \BOthers {.}}{%
Zou%
\ \protect \BOthers {.}}{%
{\protect \APACyear {2021}}%
}]{%
2021JA029117Zou}
\APACinsertmetastar {%
2021JA029117Zou}%
\begin{APACrefauthors}%
Zou, Y.%
, Walsh, B\BPBI M.%
, Shi, X.%
, Lyons, L.%
, Liu, J.%
, Angelopoulos, V.%
\BDBL {}Henderson, M\BPBI G.%
\end{APACrefauthors}%
\unskip\
\newblock
\APACrefYearMonthDay{2021}{}{}.
\newblock
{\BBOQ}\APACrefatitle {Geospace Plume and Its Impact on Dayside Magnetopause
  Reconnection Rate} {Geospace plume and its impact on dayside magnetopause
  reconnection rate}.{\BBCQ}
\newblock
\APACjournalVolNumPages{Journal of Geophysical Research: Space
  Physics}{126}{6}{e2021JA029117}.
\newblock
\begin{APACrefURL}
  \url{https://agupubs.onlinelibrary.wiley.com/doi/abs/10.1029/2021JA029117}
  \end{APACrefURL}
\newblock
\APACrefnote{e2021JA029117 2021JA029117}
\newblock
\begin{APACrefDOI} \doi{https://doi.org/10.1029/2021JA029117} \end{APACrefDOI}
\PrintBackRefs{\CurrentBib}

\end{thebibliography}

\end{document}